\documentclass[unnumsec,webpdf,contemporary,large,namedate]{oup-authoring-template}

\graphicspath{{Fig/}}

\usepackage{natbib} 
\usepackage{graphicx}
\usepackage{caption}
\usepackage{lipsum} 
\usepackage{array}
\usepackage{tabularx}
\usepackage{multirow}
\usepackage{amsmath}
\usepackage{booktabs}
\usepackage{adjustbox}
\usepackage{hyperref}
\usepackage[normalem]{ulem}

\theoremstyle{thmstyleone}

\theoremstyle{thmstyletwo}

\theoremstyle{thmstylethree}

\begin{document}

% \journaltitle{Interacting with Computers}
% \DOI{DOI HERE}
% \copyrightyear{2026}
% \pubyear{2026}
% \access{Advance Access Publication Date: Day Month Year}
% \appnotes{Paper}
\journaltitle{Interacting with Computers}
\DOI{10.1093/iwc/iwag033}
\copyrightyear{2026}
\pubyear{2026}
\access{Advance Access Publication Date: 12 June 2026}
\appnotes{Paper}

\firstpage{1}

\title[The New Social Image]{The New Social Image: How AI Competency and AI Proactivity Influence Self- and Peer-Perceptions in the Workplace}

\author[1,$\ast$]{Kuntal Ghosh}
\author[2]{Marc Hassenzahl}
\author[1]{Shadan Sadeghian}

\authormark{Ghosh et al.}

\address[1]{\orgdiv{Autonomous Interactive Systems}, \orgname{University of Siegen}, \orgaddress{\street{Kohlbettstraße 15}, \postcode{57072}, Siegen, \country{Germany}}}

\address[2]{\orgdiv{Experience \& Interaction Design}, \orgname{University of Siegen}, \orgaddress{\street{Kohlbettstraße 15}, \postcode{57072}, Siegen, \country{Germany}}}

\corresp[$\ast$]{Corresponding author. \href{mailto:kuntal.ghosh@uni-siegen.de}{kuntal.ghosh@uni-siegen.de}}

% \received{Date}{0}{Year}
% \revised{Date}{0}{Year}
% \accepted{Date}{0}{Year}
\received{16}{08}{2025}
\revised{12}{01}{2026}
\accepted{20}{04}{2026}

%\editor{Associate Editor: Name}

%\abstract{
%\textbf{Motivation:} .\\
%\textbf{Results:} .\\
%\textbf{Availability:} .\\
%\textbf{Contact:} \href{name@email.com}{name@email.com}\\
%\textbf{Supplementary information:} Supplementary data are available at \textit{Journal Name}
%online.}

\abstract
{
    Human-AI collaboration is considered the most promising way to incorporate AI in the workplace. What remains unexplored are the experiential consequences of this teaming. More specifically, in a team with AI, how humans perceive themselves (self-perception) and how they are perceived by their coworkers (peer perception) in terms of work ownership and job meaningfulness. In a $2\times2\times2$ vignette study (n=50), participants rated perceptions of ownership, affect, job meaningfulness and satisfaction, and role dynamics across two levels (low/high) of AI proactivity and AI competency as within-subject factors, with point-of-view (self perception/peer perception) as between-subjects. Our results showed that AI with low competency or low proactivity generally improved feelings related to ownership, meaningfulness, satisfaction, and role dynamics, and also increased positive affect while reducing negative affect. However, these effects were often influenced by point-of-view. For instance, low AI proactivity resulted in higher job satisfaction from self-perception rather than peer perception. Based on our findings, we argue that designing AI for the future of work solely around performance metrics may not be adequate. Highly competent and proactive AI-driven systems can have undesirable impacts on perceptions of ownership, job identity, social image and team dynamics, and consequently, job meaningfulness.
}
    

% \boxedtext{
% \begin{itemize}
% \item Key boxed text here.
% \item Key boxed text here.
% \item Key boxed text here.
% \end{itemize}}

\maketitle

\noindent\textit{This is an accepted manuscript of an article published in Interacting with Computers. The Version of Record is available at: \url{https://doi.org/10.1093/iwc/iwag033}.}

\section{Introduction}
Work is an integral part of life, with one-third of the human lifespan expended in the workplace \citep{ahmad2013paradigms}. Work fulfills basic human needs by providing survival essentials like food, shelter, and access to opportunity structures like education and healthcare systems. It also fosters social connections and a sense of community, while supporting self-determination through autonomy, competence, and empowerment \citep{duffy2016psychology, allan2020decent}. Therefore, while work provides monetary support, the quality of work life serves as a crucial source of meaning in everyday life \citep{terkel1974working, hom1994employee}. As such, a job serving a higher purpose leads to higher job meaningfulness \citep{sparks2001explaining, lips2009discriminating, laschke2020positive}. Job meaningfulness is a multi-dimensional construct that is influenced by various factors such as employees' sense of accomplishment, the stimulating challenges that foster meaningful contributions, the positive relationships they build with colleagues, the recognition they receive, and the social image they create \citep{smids2020robots, baldauf2021automation}.

Technology has been central to the advancement and transformation of work throughout history, evolving from primitive tools to modern computing devices \citep{sadeghian2024soul}.
Recently, the introduction of artificial intelligence (AI) has transformed work practices, which has led to inevitable transitions in many fields such as medicine \citep{lee2021application}, hospitality \citep{limna2023artificial}, tourism \citep{gupta2023future}, software development \citep{ebert2023generative}, and more \citep{dwivedi2021artificial, marquis2024proliferation}. 
With AI-driven systems being integrated into domains previously limited to humans (e.g., project management \citep{daugherty2018human+}), existing work practices might change \citep{bergqvist2024towards}.
Consequently, job meaningfulness could be influenced, particularly as collaborative environments shift from human-human to human-AI teams \citep{nyholm2020can}. One of the factors influencing the perception of work as meaningful lies in its social context \citep{blustein2023understanding}, specifically the value of achievement \citep{danaher2021automation} and the consequent social recognition and image \citep{gheaus2016goods}, which are reflected through acknowledgments from colleagues or organization management \citep{wrzesniewski2003interpersonal}. However, the advent of AI in the workplace can threaten these values by taking over the tasks that define one’s work identity and significance \citep{hackman1974job}, and consequently the perception of ownership over work outcomes \citep{sadeghian2024soul}.

Currently, the development of AI-driven systems for work often focuses on enhancing performance by creating models that produce more precise and less error-prone outcomes, or by designing systems that gather information from their context to make decisions and take actions proactively. However, while AI characteristics such as high proactivity can instill trust in the system \citep{kraus2020effects, kuang2024enhancing}, it can also intrude on human's decision-making space, making them feel disempowered \citep{peng2019design}. Similarly, while high AI competency can increase willingness to use the system \citep{dietvorst2015algorithm, castelo2019task}, it can also reduce the extent to which humans value the results of their work and make them feel inept \citep{danaher2021automation}. This could further influence the perception of own competency \citep{ashktorab2020human, zhang2023investigating}, role dynamics \citep{siemon2022elaborating, clear2025ai}, as well as ownership and affect \citep{kuzminykh2020mine, kobiella2024if} in human-AI teaming. Consequently, an AI with a combination of these two attributes can decrease the opportunities for humans to engage in tasks that are perceived as genuine accomplishments or receive praise and recognition.
Previous research in the field of human-AI collaboration has focused on how humans perceive AI-driven systems in general \citep{hayashi2017canai, ragot2020ai}, and to a lesser extent on their perceptions of AI as teammates in human-AI teams \citep{zhang2021ideal}. However, human self-perception and peer perception in such teams remain largely unexplored.
Since most AI-driven technologies are primarily designed to cater to the practical aspects of collaboration, such as efficiency and effectiveness, with less emphasis on fostering meaningful experiences \citep{kaasinen2015defining}, the experiential aspects of work in human-AI teams are often overlooked.
This includes how individuals perceive themselves (self-perception) and how they are perceived by their coworkers (peer perception) in the context of working with AI, and how these perceptions affect their sense of ownership, job meaningfulness, and role in the workplace.
As such, this paper aims to answer the following research questions:
\begin{itemize}
    \item \textbf{RQ1:} How do AI competency and proactivity influence self- and peer perceptions of ownership of work outcomes?
    \item \textbf{RQ2:} How do AI competency and proactivity influence self- and peer perceptions of job meaningfulness and affect in the workplace?
    \item \textbf{RQ3:} How do AI competency and proactivity influence self- and peer perceptions of role dynamics in the workplace?
\end{itemize}

To answer these questions, we conducted an experimental video vignette study, in which we asked participants to assess their self-perception and peer perception within a team meeting, where the team lead interacts with team members and with an AI with two levels of competency (low/high) and proactivity (low/high). 
Our results showed that AI with low competency or low proactivity generally improved all evaluated aspects, increasing positive affect and reducing negative affect.
However, point-of-view moderated some of these evaluations. 

The following sections start with a summary of related work on human-AI collaboration at workplaces, job meaningfulness in collaboration with AI, and proactivity in AI. We then present our study method and subsequent results, followed by a discussion of our findings. Lastly, we conclude with a reflection on future work.

\section{Related Work}

\subsection{Human-AI Collaboration at Work}
The proliferation of artificial intelligence in workplaces, often called "the fourth industrial revolution", is reshaping how businesses operate. It is pushing them to rethink their strategies, innovate in new ways, and reconsider how they measure performance \citep{longo2020value, sima2020influences, parashar2023foundation}. However, the intricate complexities of the current business landscape cannot be solved by machines alone \citep{raftopoulos2023human}. Rather, it requires a human-machine hybrid collaboration to comprehensively leverage the capabilities of both humans and AI \citep{crandall2018cooperating, akata2020research, de2021ai, dellermann2021future}. Hybrid intelligence is defined as \textit{“the ability to achieve complex goals by combining human and AI, thereby reaching superior results to those each of them could have accomplished separately, and continuously improve by learning from each other”} \citep[p. 640]{dellermann2019hybrid}. As such, a collaboration between humans and AI-driven systems combines the benefits of the human's perception, social skills, or cognitive qualities like creative problem-solving, with the advantages of AI - speed and accuracy \citep{braga2017emperor, you2018enhancing, esterwood2020human}. Leveraging these complementary skills leads to superior performance through enhanced decision-making and increased efficiency in human-AI collaboration \citep{madni2018architectural, dubey2020haco, hemmer2021human}. The positive impact of this collaboration has already been proven in the fields of education \citep{holstein2022designing}, finance \citep{lui2018artificial, buckley2021regulating}, healthcare \citep{tschandl2020human, lai2021human}, manufacturing \citep{emmanouilidis2021human}, and more.

Extensive research has explored the dynamics of human-AI collaboration in team settings.
For instance, \cite{han2024teams} found that designers often viewed AI as a tool rather than a true collaborator, questioning its ability to engage in human-like discussions or contribute effectively to decision-making. Designers consistently prioritized human input over AI suggestions, often verifying AI output with their human peers first. 
On the same note, \cite{hou2023should} discovered that while leaders, be it AI or humans, were perceived as more competent, participants preferred human leaders for their warmth and familiarity.
Such feelings of warmth also influence participants' choices between AIs of varying competencies.
\cite{gilad2021effects} found that participants preferred an AI that was friendly or helpful over one that was highly skilled, even if the more personable AI was less competent.
However, zealous AIs can significantly reduce task completion times. \cite{xu2023comparing} observed users engaged in video anonymization using two AIs - a "restrained-AI" that provided only high-precision recommendations, and a "zealous-AI" that prioritized recall through increased detections. They found that the zealous-AI enabled users to complete tasks faster with higher recall. 

The role of AI in influencing ownership and accomplishment is also significant. 
\cite{kobiella2024if} explored how AI impacts the perception of achievement and productivity. Some participants reported an increased sense of ownership, viewing the AI as an "enhancement tool" that amplified their abilities. This perception boosted their sense of accomplishment, as they felt the AI supported and built upon their original ideas. Conversely, others felt that the ease of completing tasks with minimal effort reduced their sense of achievement, as their contributions seemed limited and overshadowed by the AI's involvement.
This was complemented by \cite{xu2024makes}, who found that the originality of co-creation and the extent of control over the outcome are crucial in determining ownership and perceived contributions.

Research also highlights a preference for AIs that humans perceive as like-minded, as these AIs are seen as easier to collaborate with due to their alignment with human thinking.
\cite{guo2024exploring} investigated participants' perceptions of ownership and AI competence during brainstorming sessions, finding that AI diminished participants' sense of ownership by reducing their creative agency. However, participants found it easier to generate ideas when collaborating with AI that shared their mindset.
Similarly, \cite{gu2024data} examined how data analysts perceived AI assistants in supporting analysis planning and execution. They discovered that analysts preferred AI suggestions that aligned with their own analysis and expertise, while they were hesitant to accept suggestions that did not match their background.

In summary, while previous research on human-AI collaboration at work has explored the effects of AI characteristics such as friendliness and competence on human satisfaction and ownership of outcomes, studies examining how such AI characteristics influence self- and peer perceptions of human collaborators are still lacking.

\subsection{Job Meaningfulness in Human-AI Collaboration}
The increasing use of the term "Human–AI Collaboration" coincided with the rise of Industry 4.0, as advances in automation, efficiency, and data-driven production brought humans and intelligent systems into closer interaction at work \citep{xu2018industry}.
However, within these collaborations, emphasis was often placed on technological optimization and productivity, with comparatively less attention given to human roles, job quality, and broader societal considerations in the workplace \citep{longo2020value}.
To address these limitations, Industry 5.0 has emerged as a complementary paradigm that foregrounds human-centred values.
Discussions surrounding the transition from Industry 4.0 to Industry 5.0 emphasize a shift from purely efficiency-driven automation toward approaches that prioritize worker well-being, agency, and meaningful engagement at work \citep{alves2023industry}. 
Consequently, there has been a growing interest in exploring human-AI collaboration in the workplace beyond its performance-oriented outcomes over the years.

\cite{breazeal2004socialrobots} provided some early insights by utilizing the joint intention theory \citep{cohen1991confirmations} to propose a framework wherein a robot proactively offers help by anticipating the human's needs. 
These foundational ideas were later expanded upon by \cite{nyholm2020can} who argued that robots have the potential to be "good colleagues" and enhance job meaningfulness by being respectful, supportive, and working constructively with humans.
In line with this, \cite{walliser2019team} investigated human-AI teamwork and observed an improved affect when AI is perceived as a teammate rather than merely a tool.
This perception, as \cite{sadeghian2024soul} further discovered, directly influences job meaningfulness, where AI as a teammate enhances it, while AI as a superior diminishes it. They also highlighted the importance of balancing human autonomy and accountability in maintaining job meaningfulness.
However, \cite{smids2020robots} noted that while robots can handle repetitive tasks thereby allowing humans to focus on more intellectually stimulating work, the very presence of robots could reduce the human count at the workplace. This would lead to reduced social interactions and subsequent feelings of isolation, diminishing meaningfulness.
\cite{waardenburg2024human} further highlighted such a dual-sided outcome of automating simple customer service tasks at an organization. While expanding the emotional support aspect of the work enhanced employees' sense of meaningfulness, the transfer of simpler, "take-a-breather" tasks to the AI affected their satisfaction due to the "psychological heaviness" of their remaining work.

\cite{ficuciello2019surgicalrobots} explored meaningful human control in autonomous surgical robots. Their findings indicate that the perception of meaningful human control depends on the level of autonomy exercised by robots, and must be adaptable and tailored to contextual conditions. This concept of control is important for maintaining job meaningfulness.
Building on this idea of control, \cite{hemmer2023human} observed that AI delegation, regardless of human awareness, enhances self-efficacy and, consequently, task satisfaction. Their findings suggest that when humans think they are in control, their engagement and sense of meaningfulness in their work increase.
However, the dynamics of support in the workplace also play a significant role. \cite{ossadnik2023man} examined how the perception of organizational support influences job meaningfulness, revealing that support from human coworkers is valued more highly than support from AI. This indicates that human participation remains crucial for job meaningfulness.
While there is increasing research on job meaningfulness in human-AI collaboration, we still lack an understanding of how individuals’ self- and peer perceptions influence their perception of meaningfulness in human-AI teaming.
Addressing this gap can help us understand how different perspectives shape job meaningfulness in such teams.

\subsection{Proactivity in AI}
Advances in machine learning and AI have transformed conventional technologies from passive tools to self-sufficient, self-learning computational artifacts \citep{sadeghian2023autonomy}.
AI-driven systems can now anticipate user needs \citep{sun2016contextual}, intervene proactively \citep{pejovic2015anticipatory}, and execute tasks independently \citep{sarikaya2017technology}.

In technical systems, proactivity is defined as an \textit{“autonomous, anticipatory system-initiated behavior, with the purpose to act in advance of a future situation, rather than only reacting to it”} \citep[p. 74]{nothdurft2014finding}.
For conversational agents, proactivity is defined as \textit{"the capability to create or control the conversation by taking the initiative and anticipating impacts on themselves or human users"} \citep[p. 1]{deng2023goal}. 
So, a proactive interaction can be defined as one that is initiated by a system, based on the anticipation of what could be warranted by a target (user, observer, or self). 

\cite{meurisch2020exploring} refined the AI proactivity continuum by \cite{sarikaya2017technology} and categorized it into three classes - reactive, proactive, and autonomous.
Reactive AI systems are the most common and require user inputs (through different modalities like voice or gesture) to be of assistance. Conversational agents like Apple Siri, Amazon Alexa, Google Assistant, or Microsoft Cortana are examples of such systems that require explicit commands to enable action \citep{luger2016like, de2020intelligent}.
Proactive AI systems take preemptive actions without explicit user request. These can either constitute systems that alert users to important information, or systems that tailor their algorithmic results in line with user needs and provide recommendations to meet those needs \citep{sun2016contextual, pejovic2015anticipatory}. Modern AI-driven systems leverage predictive models to identify ideal times for sending out notifications \citep{mehrotra2017intelligent, pielot2017beyond} and intervening \citep{pejovic2015anticipatory}. However, the actual actions remain under the purview of the user. One such example is healthcare applications that provide relevant suggestions for meeting fitness goals based on the user's activity levels \citep{schmidt2015fitness, rabbi2015mybehavior}.
Autonomous AI-driven systems make decisions and take actions without needing prior confirmation from users. Some examples of such systems are digital agents \citep{yorke2012design}, social robots \citep{peng2019design}, or empirical decision-making in autonomous vehicles \citep{li2018humanlike}.

One element of proactive interaction is the AI-driven system's ability to anticipate human needs and future actions. Anticipation refers to the system’s prediction of what the human is going to do, prefer, or need in a particular situation \citep{butz2004anticipatory, cerejo2023anticipation}.
\cite{kraus2020effects, kraus2020successful} identified anticipation as either implicit or explicit. In implicit anticipation, the system takes action based on automatically gathered data, for example, through physiological measures like eye gaze \citep{billing2019proactive}, or movement \citep{gliesche2020comparison}. Contrarily, in explicit anticipation, the system takes action based on previously stated user preferences, for example, through verbal information \citep{l2005can}. Consequently, in explicit proactive interaction, the reasons behind the system's actions are more likely to be understood by the user, since they are based on explicit statements made towards the system. However, actions based on implicit data are much more difficult to understand, since they emerge from data and related behavioral patterns, which the user might not even be aware of.

Another aspect of proactive interaction is the initiation of action by AI-driven systems. 
\cite{whittaker2021designing} developed three distinct robot personas varying in proactivity -  buddy (high proactivity), butler (medium), and sidekick (low). 
They found that people prefer more proactive personas because they are perceived as more energetic and upbeat compared to less proactive ones.
\cite{peng2019design} explored three levels of proactivity (low, medium, and high) in the context of decision-making support from service robots. A robot with high proactivity made strong assumptions about user needs and either suggested appropriate actions or took them directly. On the medium level, the robot only made assumptions and let the user verify them. On a low level, the robot made no assumptions and required the user to explicitly specify their needs. They found that while a highly proactive robot offers extensive information, people find it inappropriate since it intrudes on their decision-making space. In contrast, a robot with medium proactivity is more favored as it narrows the decision space without diminishing users' sense of engagement.

\cite{meurisch2020exploring} conducted an in-the-wild study to understand user expectations of proactive AI-driven systems and the influence of different human factors on them. They found that while people are generally open to proactive support, their preferences regarding proactivity vary based on their personality, socio-demographic characteristics, and areas of support. 
Likewise, \cite{kraus2020effects} examined how different levels of proactivity (none, notification, suggestion, and intervention) along with timing strategies (fixed-timing and insecurity-based) are trusted by subjects while performing a planning task. Their results showed that proactive behavior can help render a more competent and reliable system that could ultimately lead to a more trustworthy interaction partner. However, proactive dialogue was found to significantly influence users' cognition-based trust in the system’s perceived competence and reliability. 
However, it remains unclear at what point this combination of AI competency and AI proactivity becomes too much, or too little, for the users' liking. The relationship between AI competency and AI proactivity requires further exploration to determine how each factor shapes user experiences.

\begin{figure*}[!t]
    \centering
    \includegraphics[width=\textwidth]{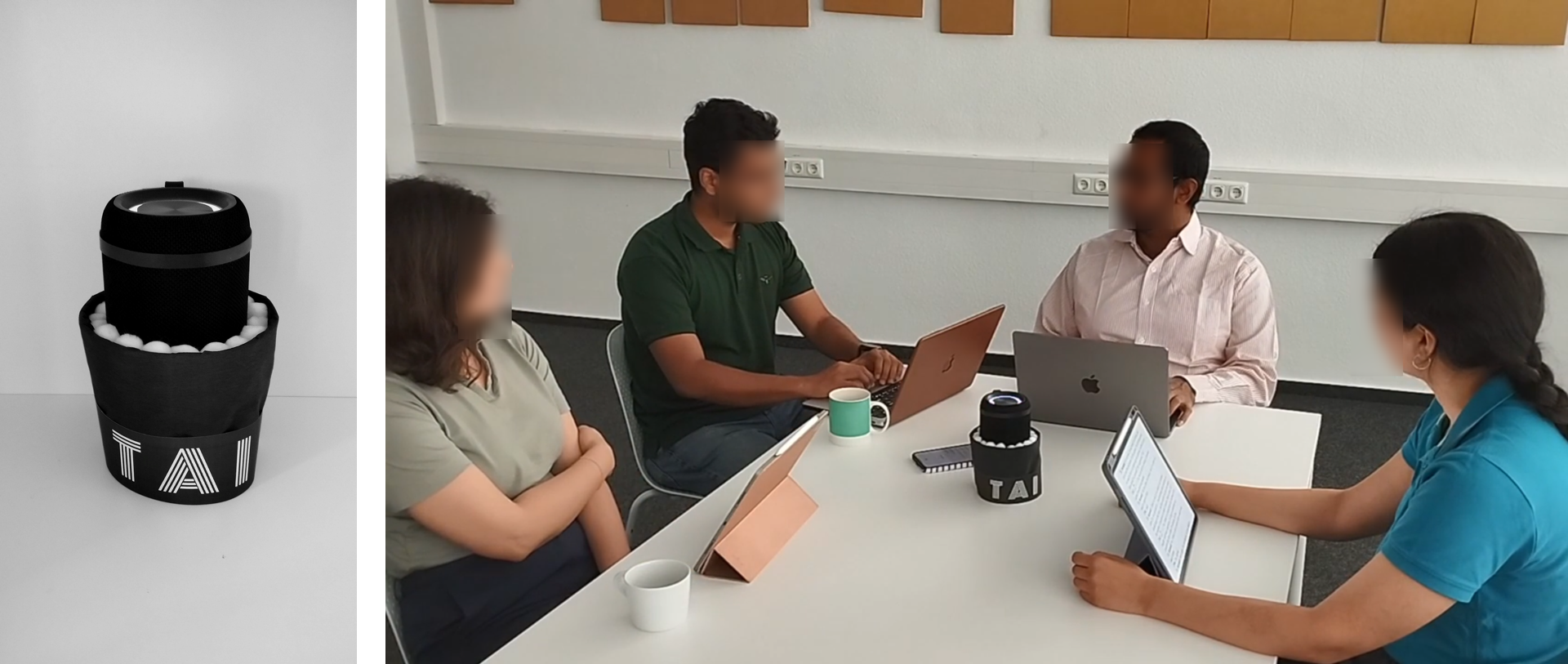}
    \caption{Left: TAI working prototype; Right: Video vignette setup showing the team lead interacting with team members and TAI.}
    \label{fig:video_vignette_setup}
\end{figure*}

\section{Method}
We conducted an online video vignette study \citep{atzmuller2010experimental, hillen2013developing, aguinis2014best} to explore how working with AI in the workplace influences self-perception and peer perception.
Experimental vignette methodology is an established approach \citep{landman1987regret, finch1987vignette, taylor2006factorial, sadeghian2022artificial, das2024sensible, neuhaus2024mimic} that enables the manipulation and control of specific variables within a controlled setting, that are often difficult to obtain through field studies. In this method, thoughtfully constructed realistic scenarios are presented to participants to measure dependent variables, including attitudes and behaviors. This enriches experimental realism and allows researchers to manipulate and control independent variables \citep{aguinis2014best}. 

Our study had a three-factorial mixed design, with the between-subject factor "point-of-view" (self, peer), and the within-subjects factors “AI Proactivity” (low, high), and “AI Competency” (low, high).
Participants were shown four videos of an in-person team meeting in an office setting, where an AI-driven system (named TAI) intervened as one team lead and three team members conversed (Figure \ref{fig:video_vignette_setup}) (see Appendix \ref{sec_appendix} and supplementary material for more details about the scenarios).
In the vignettes, TAI was positioned directly in front of the team lead on the table, symbolizing its role as the team lead’s personal collaborator. Furthermore, it directed its interventions exclusively to the team lead and did not address the other team members.
TAI was introduced in the study as follows: \textit{"The organization has introduced an AI-driven system called TAI (image provided below) to manage administrative and project-related tasks"}.

The videos were uploaded as part of an online survey created using LimeSurvey\footnote{https://www.limesurvey.org/}. This survey was distributed to participants through Prolific\footnote{https://www.prolific.com/} and administered in English. Participants completed the study individually and asynchronously, watching the videos on their own devices and responding to the survey without real-time researcher moderation.
All participants watched the four videos in which TAI varied in competency (low/high) and proactivity (low/high).
We deliberately kept the meeting context straightforward and avoided technical jargon to ensure that participants could focus on the human-AI behaviors and team dynamics, rather than on the specific content or technicalities of the meeting.
The scenarios depicted a meeting involving typical office administrative topics, such as planning vacations or discussing project updates, designed to be simple yet realistic and familiar to participants across professional roles.
For example, during the team meeting in the vignettes, the team lead asked the team members to share their vacation plans.
When TAI was highly competent and proactive, it took the initiative by suggesting an efficient solution: \textit{“Everyone can simply mark it on their digital calendars … you can see the entire plan at a glance, and also notice if there are overlaps!”} In contrast, when TAI had low competency and low proactivity, it responded only after being asked, with a less efficient approach: \textit{“Everyone can send you an email regarding their vacation plans, and then you can decide.”}
The order of the four videos was randomized, with no written text specifying the level of competency and proactivity of TAI.
Half of the participants watched the videos and answered questions from the perspective of team members, focusing on how they perceived the team lead (peer perception). The other half watched the same videos but responded from the perspective of the team lead (self-perception).
This allowed us to independently analyze the effects of human-AI teaming on each type of perception without carryover effects between responses.

\subsection{Design of TAI}
The TAI prototype was designed to simulate an AI-driven interactive system capable of conversing through speech.
The objective was to integrate TAI into a team meeting with human members and assess how its behavior (based upon competency and proactivity) influenced the team lead's self-perception and how he was perceived by his peers (peer perception).

The TAI prototype consisted of a Bluetooth speaker, fitted within an earthen pot padded with cotton balls, which served as the base of the prototype. The pot was camouflaged with crepe paper, and prominently displayed the prototype's name "TAI". 
To simulate AI-generated speech, audio was produced, in advance, using TTSMP3\footnote{https://ttsmp3.com/}. 
We selected an AI voice named Nova, characterized by a high-pitched, soft, and friendly tone with an American English accent.
These audio clips were manually triggered during the recording of the video vignettes using a computer, ensuring alignment with the scripted interactions requiring TAI's involvement.

Based on the AI proactivity categorization from \cite{meurisch2020exploring}, we designed TAI to be "reactive" in two vignettes (where it had low proactivity and responded only when prompted by its name) and "proactive" in two vignettes (when it had high proactivity, and intervened frequently with its comments or opinion).
Across all four vignettes, TAI's interactions were designed on the principles of implicit anticipation by \cite{kraus2020effects, kraus2020successful}, wherein it replied using insights derived from the meeting discussion.
When AI competency was low, TAI gave vague, partially incorrect, or unhelpful suggestions that did not fully align with the discussion context. When AI competency was high, TAI gave precise, contextually relevant, and well-informed answers.

\subsection{Measures}
Our questionnaire had three sections for each video vignette. However, before asking questions about our dependent variables, we needed to validate the levels of our independent variables in the scenarios. Therefore, after participants watched each video, we asked them to rate the competency and proactivity of TAI on a 7-point Likert scale (1=Low, 7=High), and how well they could imagine themselves in the presented situations on another 7-point Likert scale (1=Not at all, 7=Extremely well).

\subsubsection{Ownership}
The first part involved understanding how participants perceived human-AI ownership over work processes and outcomes across the scenarios. We utilized the Psychological Ownership Questionnaire (POQ) by \cite{avey2009psychological} which is used to understand the experiences arising due to various positive (“promotive”) and negative (“preventative”) psychological states at work. The original questionnaire comprises five sub-scales: Territoriality (4 items): reluctance to share resources or information, or being in control of what to share with whom, Self-efficacy (3 items): belief in one's ability to successfully perform tasks, Accountability (3 items): sense of duty and responsibility, Belongingness (3 items): extent to which employees feel at home in their workplace, and Self-Identity (3 items): identifying with the organization as an extension of themselves. While Territoriality is characterized as preventative psychological ownership, the remaining four sub-scales are considered as promotive. For our study, we customized and used an altered form of the POQ (Cronbach’s $\alpha$ = .89) after obtaining permission from the publisher, Mind Garden, Inc\footnote{https://www.mindgarden.com/}. We used three out of five sub-scales, each consisting of 3 items - Territoriality (Cronbach’s $\alpha$ = .77), Self-Efficacy (Cronbach’s $\alpha$ = .89), and Accountability (Cronbach’s $\alpha$ = .80), all reporting an acceptable level of reliability \citep{ursachi2015reliable}. The nine items were rated on a 7-point Likert Scale (1 = Strongly disagree, 7 = Strongly agree).

\subsubsection{Job Meaningfulness and Affect}
To measure job meaningfulness and satisfaction, we asked participants to fill in the Job Diagnostic Survey (JDS) by \cite{hackman1974job, hackman1975development}. It includes five sub-scales; Skill Variety (SV) (5 items, Cronbach’s $\alpha$ = .87): the extent to which employees get to learn and use a wide variety of skills, Task Identity (TI) (4 items, Cronbach’s $\alpha$ = .73): the extent to which the job involves completing distinct tasks with clear outcomes, Task Significance (TS) (4 items, Cronbach’s $\alpha$ = .82): the extent to which employees find their job substantial and impacting the lives of others, Autonomy (AU) (4 items, Cronbach’s $\alpha$ = .77): the extent to which employees have control in making decisions and taking actions, and Feedback (FB) (6 items, out of which we used 2, Cronbach’s $\alpha$ = .80): the extent to which employees receive information regarding their performance. As such, we customized the JDS to fit our study (Cronbach’s $\alpha$ = .93). The individual sub-scales are then used to calculate an overall Motivating Potential Score (MPS) by using this formula: (\( \text{MPS} = \frac{\text{SV} + \text{TI} + \text{TS}}{3} \times \text{AU} \times \text{FB} \)) \citep{hackman1974job}.  All items were rated on a 5-point Likert scale (1 = Strongly disagree, 5 = Strongly agree). We included all five sub-scales in our analysis since all had an acceptable level of reliability \citep{ursachi2015reliable}.

Since JDS does not directly inquire about the perception of job meaningfulness and satisfaction, we asked participants to rate their perception of meaningfulness and satisfaction.
Participants answered these two questions on a 7-point Likert scale and described why they felt so in additional open questions. Moreover, we asked participants how positive/negative they found the situations portrayed in the video on a 7-point Likert scale, and their reasons behind the positive/negative affect (open question).

\subsubsection{Role Dynamics}
To explore how the relationship with AI was influenced by AI competency and AI proactivity from the point-of-view of self and peers, we asked participants to rate TAI as a superior/subordinate/teammate across the four scenarios on a 7-point Likert scale. We also asked them about the reasons behind their corresponding ratings (open question).

\subsection{Participants}
To determine the required sample size for our study, we conducted an a priori power analysis using G*Power\footnote{https://www.psychologie.hhu.de/arbeitsgruppen/allgemeine-psychologie-und-arbeitspsychologie/gpower}. Based on guidelines by \cite{cohen2013statistical}, we assumed a medium effect size (f=0.25), an alpha level of 0.05, and a power of 0.80. The analysis indicated that a minimum of 24 participants per point-of-view was required to achieve the desired power of 0.80.
As such, we recruited 50 participants through Prolific - 25 for each point-of-view (team lead and team member), which met the recommended sample size and ensured adequate statistical power for detecting significant effects.
25 participants (13 female, 12 male, 0 diverse) aged between 26-51 years (M=36.16, SD=8.07) answered the questions from the point-of-view of the team members (peer perception), and another 25 participants (14 female, 11 male, 0 diverse) aged between 21-56 years (M=37.84, SD=10.42), answered from the perspective of the team lead (self-perception). 

To increase the validity of our study, we filtered participants based on their professional roles using Prolific’s pre-screening feature. Participants who reported to a supervisor were placed in the peer perception group, where they responded as team members. Conversely, participants in supervisory or leadership roles were assigned to the self-perception group, where they responded as the team lead. This pre-screening process ensured that participants’ real-world experiences aligned with the perspective they were asked to evaluate. 
Prolific required all the participants and us to agree to their terms of use, including ethical data handling guidelines. Furthermore, before commencing the study, participants were informed that the collected data would be used solely for research purposes, with no individually identifiable information being disclosed. The survey was designed to take approximately 45 minutes to complete, and participants were compensated at £8.00/hour.

\section{Results}
To verify if participants perceived the varying levels of AI proactivity and AI competency in our video vignettes as they were meant to, we asked them to rate the proactivity and competency of TAI for each scenario (low competency-low proactivity (LCLP); low competency-high proactivity (LCHP); high competency-low proactivity (HCLP); high competency-high proactivity (HCHP)). 
We conducted paired samples t-tests to compare participants' responses across different conditions.
The results showed significant differences between low and high competency vignettes (\(t_{\text{LCLP-HCLP}}(49)=-10.14,\;p<0.001\); \(t_{\text{LCHP-HCHP}}(49)=-11.44,\;p<0.001\)), and between low and high proactivity vignettes (\(t_{\text{LCLP-LCHP}}(49)=-3.01,\;p=0.004\); \(t_{\text{HCLP-HCHP}}(49)=-7.73,\;p<0.001\)). 
This implies that participants perceived the levels of AI competency and AI proactivity as intended across our four vignettes.

Participants could also imagine themselves well in the four scenarios 
(LCLP: M=4.46, SD= 1.89; LCHP: M=3.84, SD=1.98; HCLP: M=5.44, SD= 1.53; HCHP: M=4.04, SD=1.95), with all ratings above the midpoint of the 7-point Likert scale. 
The lowest ratings were for the scenario where AI competency was low and AI proactivity was high. Participants found it harder to imagine themselves in this scenario compared to the others due to TAI's \textit{"annoying"} behavior, rather than their perceptions of the vignette. 
In the following, we present our results according to our research questions.

\begin{table*}[t]
\centering
\begin{tabularx}{\textwidth}{|l|c|XX|XX|XX|XX|}
\hline
\textbf{Scale} & \textbf{Point-of-view} & \multicolumn{2}{c|}{\textbf{LCLP}} & \multicolumn{2}{c|}{\textbf{LCHP}} & \multicolumn{2}{c|}{\textbf{HCLP}} & \multicolumn{2}{c|}{\textbf{HCHP}} \\  
 & & \textbf{Mean} & \textbf{SD} & \textbf{Mean} & \textbf{SD} & \textbf{Mean} & \textbf{SD} & \textbf{Mean} & \textbf{SD} \\ \hline
\multirow{2}{*}{\textbf{Territoriality}} & Self & 5.32 & 1.59 & 3.96 & 1.95 & 5.48 & 1.36 & 2.19 & 0.96 \\  
 & Peer & 5.39 & 1.35 & 4.45 & 1.47 & 5.72 & 1.14 & 2.76 & 1.38 \\ \hline
\multirow{2}{*}{\textbf{Self-Efficacy}} & Self & 4.55 & 1.86 & 3.87 & 1.89 & 5.79 & 1.09 & 4.21 & 1.58 \\  
 & Peer & 5.05 & 1.44 & 4.73 & 1.48 & 5.61 & 1.12 & 4.59 & 1.55 \\ \hline
\multirow{2}{*}{\textbf{Accountability}} & Self & 5.41 & 1.25 & 5.45 & 0.97 & 5.72 & 1.00 & 4.53 & 1.52 \\  
 & Peer & 5.92 & 1.03 & 6.00 & 0.98 & 5.37 & 1.31 & 4.16 & 1.32 \\ \hline
\end{tabularx}
\caption{Mean and standard deviations of Ownership scales.}
\label{table:ownership}
\end{table*}

\subsection{RQ1: How do AI competency and proactivity influence self- and peer perceptions of ownership in the workplace?}
Ownership was assessed through three sub-scales of the customized Psychological Ownership Questionnaire: Territoriality (reluctance to share resources or information), Self-Efficacy (belief in one’s ability to perform tasks successfully), and Accountability (sense of duty and responsibility).
For each scenario, participants rated the team lead's sense of ownership while interacting with TAI on these three sub-scales, responding from either the team lead’s perspective (self-perception) or team member's perspective (peer perception).

Table \ref{table:ownership} indicates the descriptive statistics for each scale of the customized psychological ownership questionnaire.
We conducted three $2\times2\times2$ ANOVA with AI competency (low, high) and AI proactivity (low, high) as within-subject and point-of-view (self, peer) as between-subject factors.
We further conducted Tukey’s HSD tests for post-hoc comparisons following significant main effects or interaction effects.

\begin{figure}[!htbp]
    \centering
    \begin{minipage}[b]{0.48\textwidth}
        \centering
        \includegraphics[width=\linewidth]{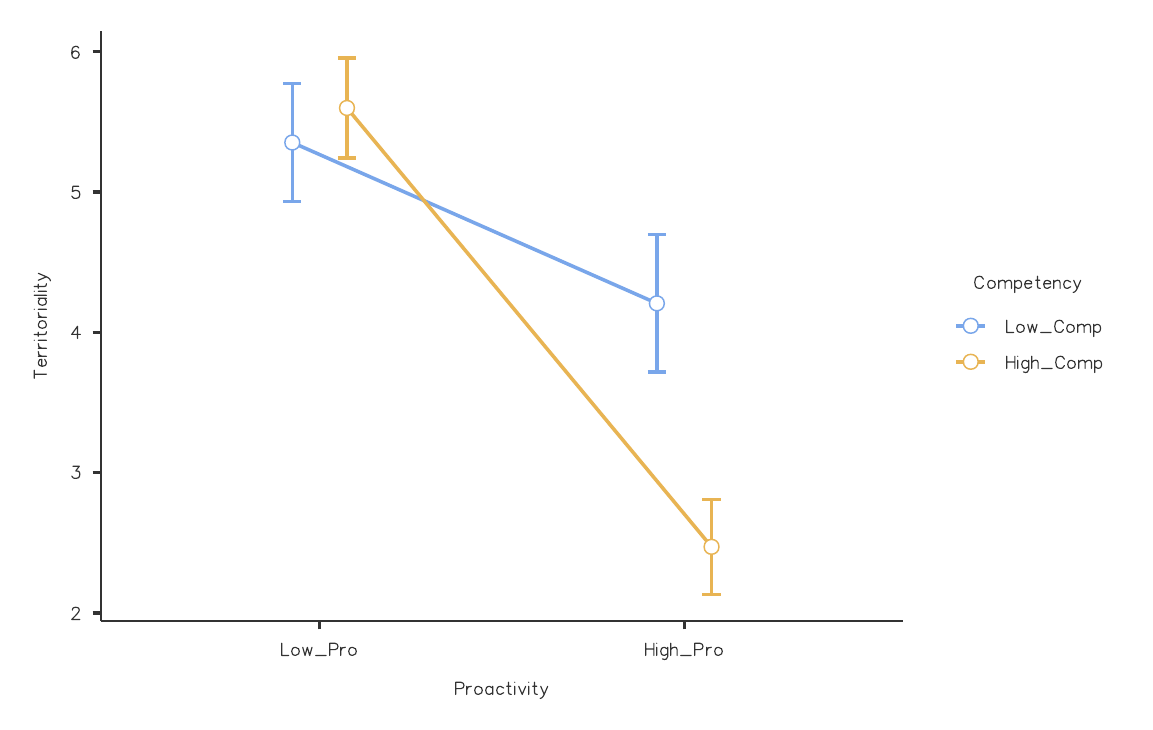}
        \phantomsection
        \label{fig:own_terr_3}
        \textbf{(a)}~Feelings of Territoriality
    \end{minipage}
    \hfill
    \begin{minipage}[b]{0.48\textwidth}
        \centering
        \includegraphics[width=\linewidth]{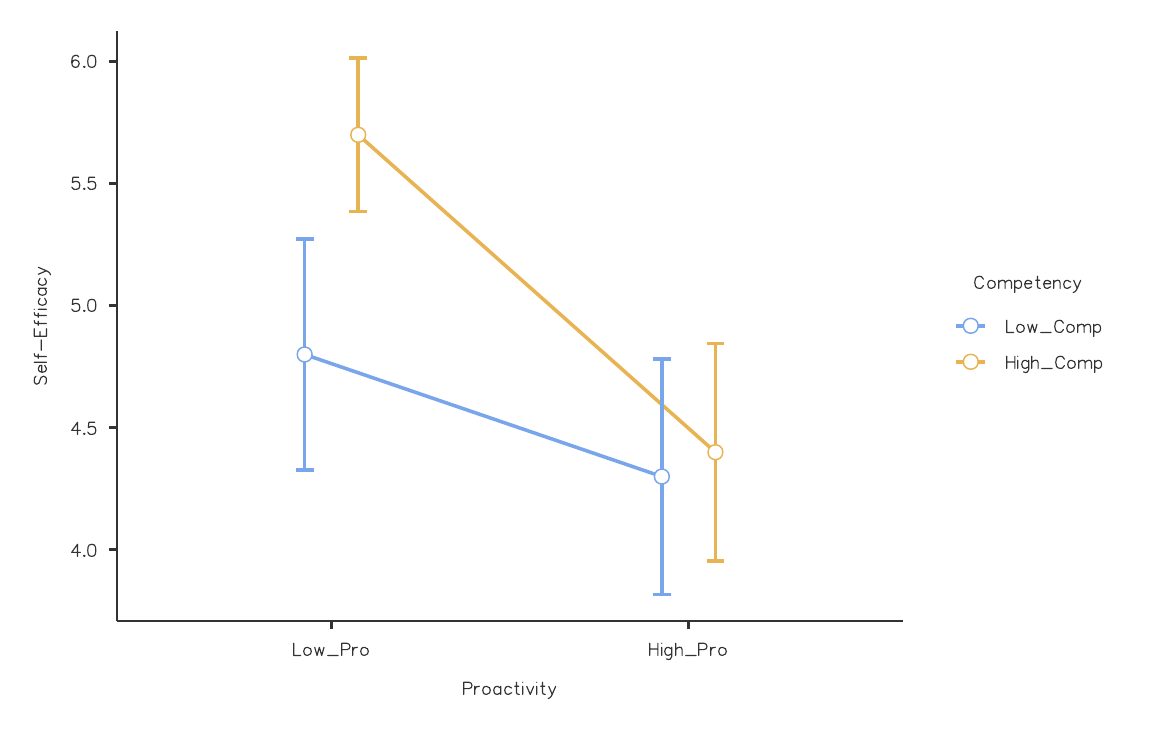}
        \phantomsection
        \label{fig:own_eff_3}
        \textbf{(b)}~Feelings of Self-Efficacy
    \end{minipage}  
    \caption{Means and 95\% confidence intervals of the customized Ownership Questionnaire for Territoriality and Self-Efficacy across low/high AI competency and AI proactivity.}
    \label{fig:own_posthoc}
\end{figure}

For the Territoriality sub-scale, we found a significant main effect of competency (\(F(1,48)=14.37, p<0.001, \eta^2=0.039\)), wherein low AI competency resulted in higher feelings of territoriality compared to high AI competency (\(t(48)=3.79,\;p<0.001\)). Similarly, a significant main effect of proactivity (\(F(1,48)=122.96, p<0.001, \eta^2=0.324\)) indicated low AI proactivity leading to higher feelings of territoriality compared to high AI proactivity (\(t(48)=11.1,\;p<0.001\)).
A significant interaction effect between competency and proactivity was observed (\( F(1,48)=46.60, p<0.001, \eta^2=0.070\)). 
Participants generally reported greater territoriality when AI proactivity was low compared to high. Notably, this effect was more pronounced when AI competency was high (\(t(48)=6.24,\;p<0.001\)) (Figure~\ref{fig:own_posthoc}a).

Meanwhile, for the Self-Efficacy sub-scale, we found a significant main effect of AI competency (\(F(1,48)=5.98, p=0.018, \eta^2=0.024\)), wherein high AI competency resulted in higher feelings of self-efficacy  (\(t(48)=-2.45,\;p<0.018\)). Similarly, a significant main effect of proactivity (\(F(1,48)=27.88, p<0.001, \eta^2=0.077\)) indicated that participants felt increased capability and confidence in their ability to perform tasks effectively when TAI had low proactivity compared to high proactivity (\(t(48)=5.28,\;p<0.001\)).
A significant interaction effect between competency and proactivity was observed (\( F(1,48)=11.74, p=0.001, \eta^2=0.015\)). Overall, participants perceived higher self-efficacy when AI proactivity was low compared to high. However, this difference was bigger when AI competency was high (\(t(48)=-4.07,\;p<0.001\)) (Figure~\ref{fig:own_posthoc}b).

\begin{figure*}[!htbp]
    \centering
    \begin{minipage}[b]{0.48\textwidth}
        \centering
        \includegraphics[width=\linewidth]{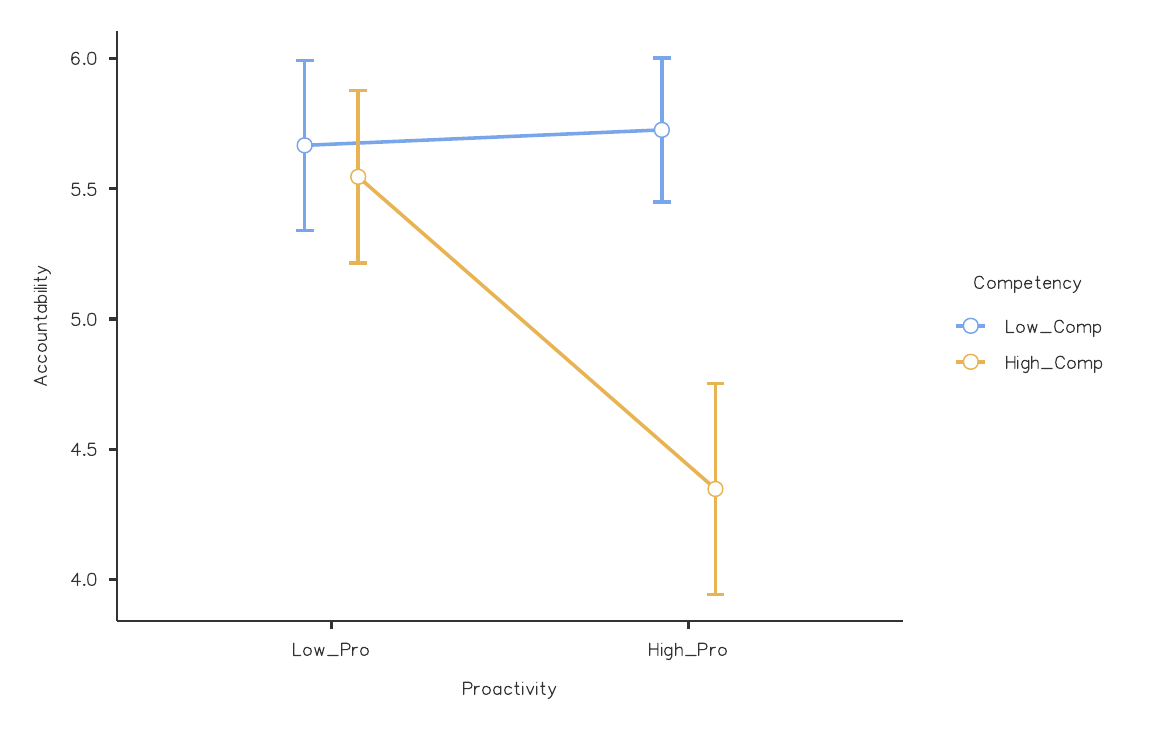}
        \phantomsection
        \label{fig:own_acc_3}
        \textbf{(a)}~Interaction effect between AI competency and AI proactivity on feelings of accountability
    \end{minipage}
    \hfill
    \begin{minipage}[b]{0.48\textwidth}
        \centering
        \includegraphics[width=\linewidth]{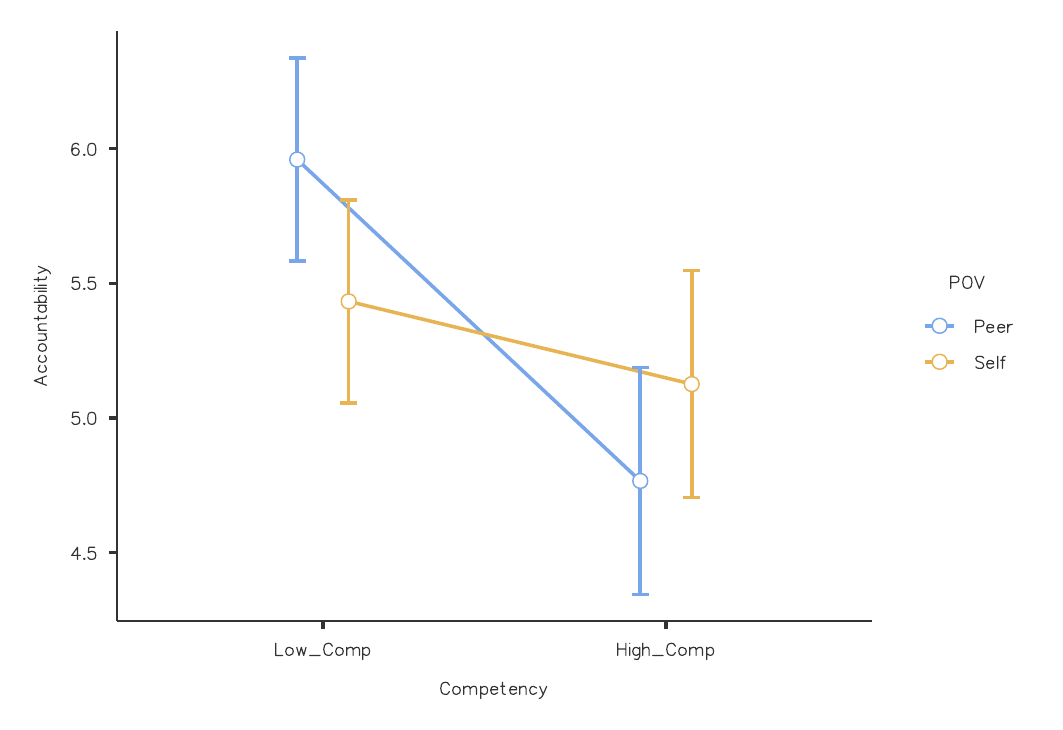}
        \phantomsection
        \label{fig:own_acc_2}
        \textbf{(b)}~Interaction effect between AI competency and point-of-view on feelings of accountability
    \end{minipage}  
    \caption{Means and 95\% confidence intervals of the customized Ownership Questionnaire for Accountability across low/high AI competency and AI proactivity, or point-of-view.}
    \label{fig:own_posthoc_acc}
\end{figure*}

Lastly, the Accountability sub-scale too showed significant main effects of AI competency (\(F(1,48)=29.62, p<0.001, \eta^2=0.081\)), wherein humans perceived higher accountability when TAI had low competency compared to high competency (\(t(48)=5.44,\;p<0.001\)). Similarly, a significant main effect of proactivity (\(F(1,48)=16.88, p<0.001, \eta^2=0.047\)) showed higher accountability when AI proactivity was low compared to high (\(t(48)=4.11,\;p<0.001\)).
Again, a significant interaction effect between competency and proactivity was observed (\( F(1,48)=27.12, p<0.001, \eta^2=0.057\)). 
Although accountability increased when AI proactivity was low compared to high, it was only evident when AI competency was high (\(t(48)=7.49,\;p<0.001\)) (Figure~\ref{fig:own_posthoc_acc}a).
Furthermore, a significant interaction effect was also observed between competency and point-of-view (\( F(1,48)=10.34, p=0.002, \eta^2=0.028\)).
In general, participants perceived higher accountability when AI competency was low compared to high, with this difference being more substantial for peer perception than for self-perception (\(t(48)=6.12,\;p<0.001\)) (Figure~\ref{fig:own_posthoc_acc}b).

\begin{table*}[b]
\centering
\begin{tabularx}{\textwidth}{|l|c|XX|XX|XX|XX|}
\hline
\textbf{Scale} & \textbf{Point-of-view} & \multicolumn{2}{c|}{\textbf{LCLP}} & \multicolumn{2}{c|}{\textbf{LCHP}} & \multicolumn{2}{c|}{\textbf{HCLP}} & \multicolumn{2}{c|}{\textbf{HCHP}} \\ 
 & & \textbf{Mean} & \textbf{SD} & \textbf{Mean} & \textbf{SD} & \textbf{Mean} & \textbf{SD} & \textbf{Mean} & \textbf{SD} \\ \hline
 \multirow{2}{*}{\textbf{Motivating Potential Score}} & Self & 46.03 & 18.27 & 43.90 & 25.88 & 50.21 & 21.89 & 21.54 & 17.63 \\ 
 & Peer & 48.27 & 27.46 & 41.06 & 21.99 & 40.77 & 21.90 & 18.12 & 16.50 \\ \hline
\multirow{2}{*}{\textbf{Skill Variety}} & Self & 3.51 & 0.58 & 3.34 & 0.70 & 3.47 & 0.60 & 2.64 & 0.84 \\  
 & Peer & 3.44 & 0.82 & 3.33 & 0.85 & 3.24 & 0.88 & 2.39 & 0.82 \\ \hline
\multirow{2}{*}{\textbf{Task Identity}} & Self & 3.39 & 0.78 & 3.11 & 0.74 & 3.41 & 0.66 & 2.83 & 0.75 \\ 
 & Peer & 3.50 & 0.85 & 3.29 & 0.83 & 3.43 & 0.74 & 2.35 & 0.72 \\ \hline
\multirow{2}{*}{\textbf{Task Significance}} & Self & 3.62 & 0.59 & 3.71 & 0.74 & 3.45 & 0.80 & 2.90 & 0.95 \\  
 & Peer & 3.81 & 0.77 & 3.84 & 0.80 & 3.57 & 0.92 & 2.76 & 0.88 \\ \hline
\multirow{2}{*}{\textbf{Autonomy}} & Self & 3.76 & 0.73 & 3.73 & 0.69 & 3.71 & 0.76 & 2.44 & 0.76 \\  
 & Peer & 3.99 & 0.94 & 3.89 & 0.94 & 3.59 & 0.86 & 2.40 & 0.92 \\ \hline
\multirow{2}{*}{\textbf{Feedback}} & Self & 3.47 & 0.88 & 3.24 & 1.08 & 3.81 & 0.78 & 2.57 & 1.14 \\ 
 & Peer & 3.17 & 1.05 & 2.93 & 0.82 & 3.14 & 1.00 & 2.44 & 0.91 \\ \hline
\end{tabularx}
\caption{Mean and standard deviations of JDS ratings.}
\label{table:jds_mean}
\end{table*}

\subsection{RQ2: How do AI competency and proactivity influence self- and peer perceptions of job meaningfulness and affect in the workplace?}

\subsubsection{Job Diagnostic Survey (JDS)}
\begin{figure}[!htbp]
    \centering
    \includegraphics[width=1.0\linewidth]{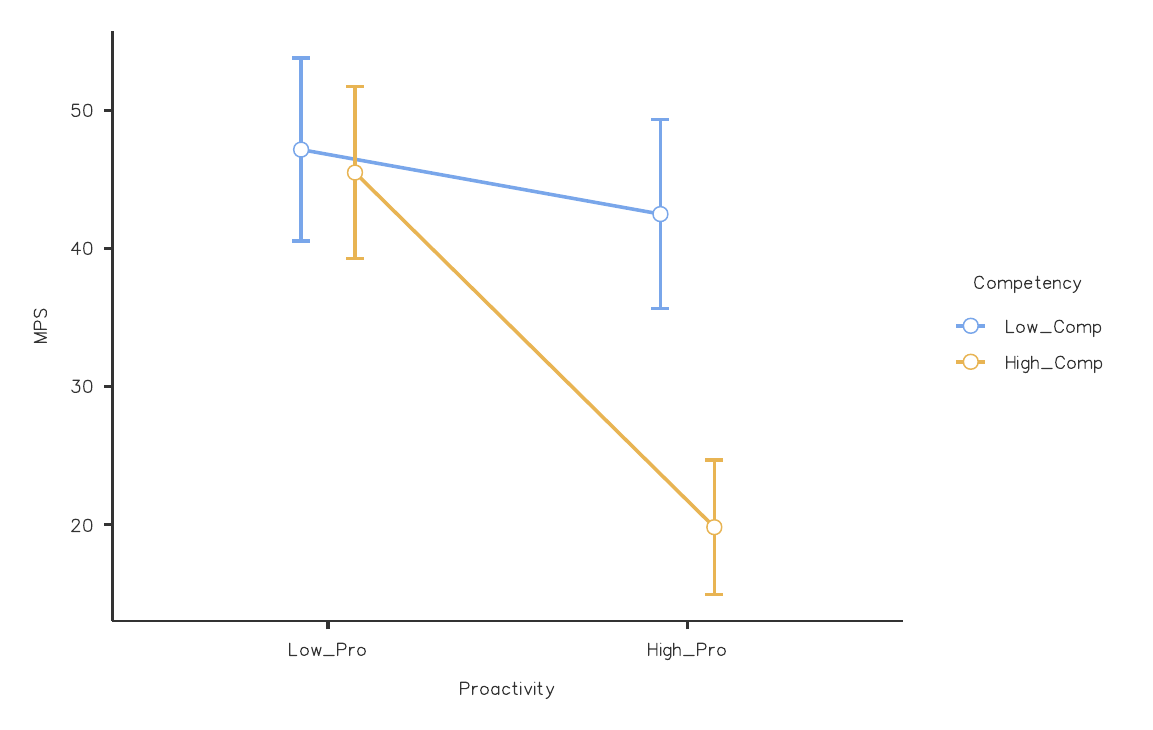}
    \caption{Means and 95\% confidence intervals of the job diagnostic survey overall motivating potential scores (MPS) across low/high AI competency and AI proactivity.}
    \label{fig:jds_mps_3} 
\end{figure}

Table~\ref{table:jds_mean} indicates the descriptive statistics for the Motivating Potential Score (MPS) of the JDS and each scale. 
Results from a $2\times2\times2$ ANOVA for MPS revealed a significant main effect of AI competency (\(F(1,48)=22.69, p<0.001, \eta^2=0.063\)), with a subsequent post-hoc test showing participants exhibited higher work motivation when TAI had low competency compared to high competency (\( t(48)=4.76, \; p<0.001 \)). Similarly, a significant main effect of proactivity (\(F(1,48)=63.85, p<0.001, \eta^2=0.099\)) indicated low AI proactivity increased work motivation while high AI proactivity decreased it (\(t(48)=7.99,\;p<0.001\)).
A significant interaction effect between competency and proactivity was observed (\( F(1,48)=27.54, p<0.001, \eta^2=0.047\)). Participants generally felt more motivated to work when TAI was less proactive. However, this drop in motivation with high proactivity was much stronger when TAI was also highly competent (\(t(48)=6.22,\;p<0.001\)) (Figure \ref{fig:jds_mps_3}).

By conducting five similar $2\times2\times2$ ANOVAs, we analyzed the differences in each scale of JDS. Similar to the results from overall MPS, we found a significant main effect of competency and proactivity for Skill Variety, Task Identity, Task Significance, and Autonomy (Table \ref{table:jds_anova_results}).
Post-hoc comparisons revealed that participants felt their jobs would allow them to use a variety of skills, be involved in tasks from start to finish, create impact, and make their own decisions when TAI had low competency compared to high, or low proactivity compared to high (Table \ref{table:jds_posthoc_results}).
Interestingly, we observed a significant main effect of only proactivity on Feedback ratings (\( F(1,48)=43.91, p<0.001, \eta^2=0.085\)), indicating that participants felt they would receive better feedback about their performance from coworkers, and from the job itself, when TAI had low proactivity compared to high proactivity (\(t(48)=6.63,\;p<0.001\)).
No significant main effect of point-of-view (self/peer) on any of the JDS sub-scales was observed.

A significant interaction effect between competency and proactivity was also observed for Task Significance (\( F(1,48)=22.81, p<0.001, \eta^2=0.043\)).
While participants perceived greater task significance at low AI proactivity level, the effect was only apparent when AI competency was high (\(t(48)=6.45,\;p<0.001\)).
Significant interaction effect between competency and proactivity was also observed for Skill Variety (\( F(1,48)=25.51, p<0.001, \eta^2=0.042\)), Autonomy (\( F(1,48)=36.24, p<0.001, \eta^2=0.083\)), Task Identity (\( F(1,48)=10.03, p=0.003, \eta^2=0.030\)), and Feedback (\( F(1,48)=10.34, p=0.002, \eta^2=0.032\)).
Participants tended to perceive higher skill variety, autonomy, task identity, and feedback at low AI proactivity level, especially when AI competency was high (\(t_{\text{SV}}(48)=6.57,\;p_{\text{SV}}<0.001\); \(t_{\text{AU}}(48)=8.31,\;p_{\text{AU}}<0.001\)); \(t_{\text{TI}}(48)=4.44,\;p_{\text{TI}}<0.001\)); \(t_{\text{FB}}(48)=3.35,\;p_{\text{FB}}=0.008\)).
There was also a significant interaction effect between competency and point-of-view for Task Identity (\( F(1,48)=4.51, p=0.039, \eta^2=0.013\)). Generally, participants reported having higher task identity (completing distinct tasks with clear outcomes) when AI competency was low compared to high. However, this difference was bigger for peer perception compared to self (\(t(48)=4.03,\;p=0.001\)).

\subsubsection{Job Meaningfulness and Satisfaction}
\begin{figure}[!htbp]
    \centering
    \includegraphics[width=1.0\linewidth]{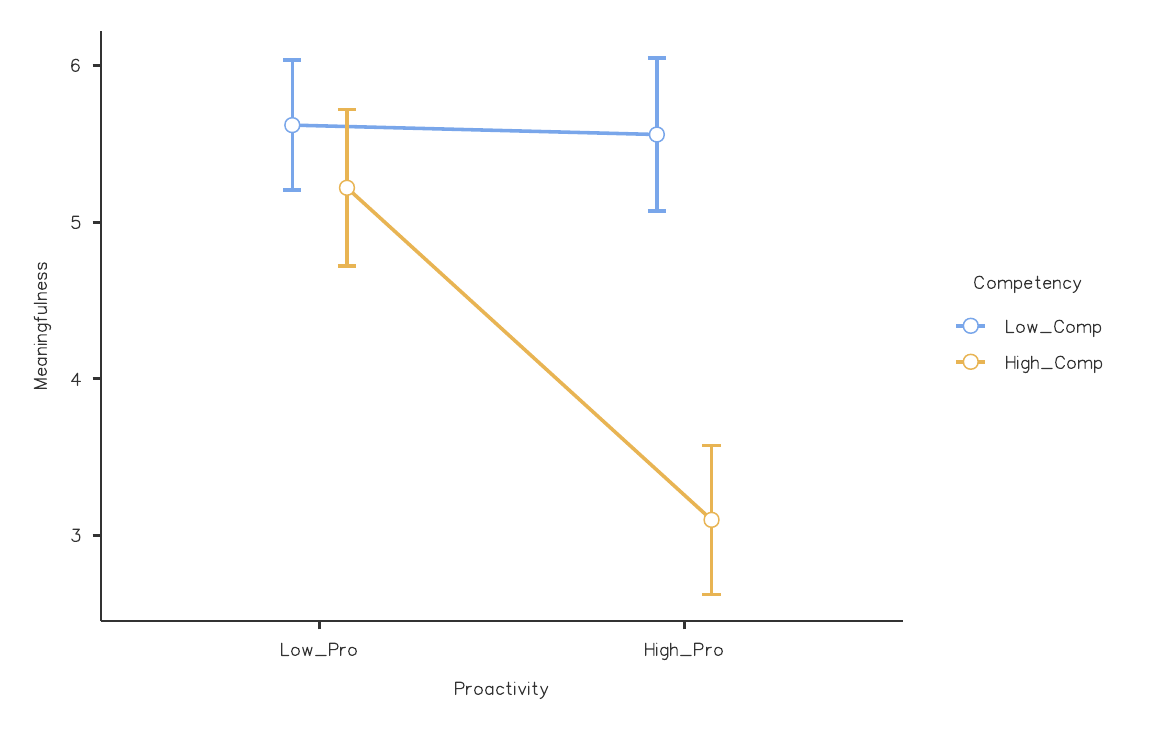}
    \caption{Means and 95\% confidence intervals of job meaningfulness across low/high AI competency and AI proactivity.}
    \label{fig:meaning_posthoc} 
\end{figure}

We asked participants how meaningful they found the job in each scenario.
The results of a $2\times2\times2$ ANOVA for meaningfulness showed significant main effects of competency (\(F(1,48)=41.81, p<0.001, \eta^2=0.136\)) wherein participants exhibited higher perception of meaningfulness when TAI had low competency compared to high competency (\(t(48)=6.47,\;p<0.001\)).
Similarly, a significant main effect of proactivity (\(F(1,48)=33.89, p<0.001, \eta^2=0.079\)) indicated low AI proactivity increased job meaningfulness while high AI proactivity decreased it (\(t(48)=5.82,\;p<0.001\)).
A significant interaction effect between competency and proactivity was observed (\( F(1,48)=43.08, p<0.001, \eta^2=0.070\)). Generally, participants felt they had higher meaningfulness when AI proactivity was low compared to high. However, this difference was only visible when AI competency was high (\(t(48)=8.85,\;p<0.001\)) (Figure~\ref{fig:meaning_posthoc}).

\begin{table}[t]
\small
\centering
\begin{tabular}{|l|p{1.9cm}|c|c|c|c|}
\hline
\textbf{Sub-scale} & \textbf{Independent Variables} & \textbf{F-statistic} & \textbf{df} & \textbf{p-value} & \textbf{\(\eta^2\)} \\ \hline
\multirow{2}{*}{Skill Variety} 
    & Competency & 30.54 & (1,48) & <0.001 & 0.076 \\ 
    & Proactivity & 39.13 & (1,48) & <0.001 & 0.083 \\ \hline
\multirow{2}{*}{Task Identity} 
    & Competency & 12.81 & (1,48) & <0.001 & 0.036 \\ 
    & Proactivity & 47.79 & (1,48) & <0.001 & 0.105 \\ \hline
\multirow{2}{*}{Task Significance} 
    & Competency & 31.78 & (1,48) & <0.001 & 0.106 \\ 
    & Proactivity & 18.54 & (1,48) & <0.001 & 0.031 \\ \hline
\multirow{2}{*}{Autonomy} 
    & Competency & 55.69 & (1,48) & <0.001 & 0.159 \\ 
    & Proactivity & 56.08 & (1,48) & <0.001 & 0.103 \\ \hline
\end{tabular}
\caption{ANOVA results for four JDS sub-scales by AI Competency and AI Proactivity.}
\label{table:jds_anova_results}
\end{table}

\begin{table}[t]
\small
\centering
\begin{tabular}{|l|l|c|c|c|}
\hline
\textbf{Sub-scale} & \textbf{Independent Variables} & \textbf{t-statistic} & \textbf{df} & \textbf{p-value} \\ \hline
\multirow{2}{*}{\textbf{Skill Variety}}   
    & Competency  & 5.53 & 48 & <0.001 \\ 
    & Proactivity & 6.26 & 48 & <0.001 \\ \hline
\multirow{2}{*}{\textbf{Task Identity}}
    & Competency  & 3.58 & 48 & <0.001 \\ 
    & Proactivity & 6.91 & 48 & <0.001 \\ \hline
\multirow{2}{*}{\textbf{Task Significance}}
    & Competency  & 5.64 & 48 & <0.001 \\ 
    & Proactivity & 4.31 & 48 & <0.001 \\ \hline
\multirow{2}{*}{\textbf{Autonomy}}
    & Competency  & 7.46 & 48 & <0.001 \\ 
    & Proactivity & 7.49 & 48 & <0.001 \\ \hline
\end{tabular}
\caption{Post-hoc test results for four JDS sub-scales by AI Competency and AI Proactivity.}
\label{table:jds_posthoc_results}
\end{table}

These ratings can be elaborated by the results of the open question which asked participants the reasons behind their ratings. 
With low AI competency-high AI proactivity, participants felt the team lead was \textit{"the final decision maker"}, who was \textit{"in charge of the room"}, and \textit{"[controlled] the outcome of the meeting"}. He was seen as not only challenging TAI, but also \textit{"educating TAI [and] the team"}, which brought meaningfulness to his job.
However, with high AI competency-high AI proactivity, participants felt the team lead's role became redundant and that he was then "TAI's assistant". Participants mentioned that \textit{"TAI took [away] the whole glory"} by constantly \textit{"taking over and not allowing [the team lead] to express [his] thoughts and contribute anything meaningful"}.

\begin{figure*}[!htbp]
    \centering
    \begin{minipage}[b]{0.48\textwidth}
        \centering
        \includegraphics[width=\textwidth]{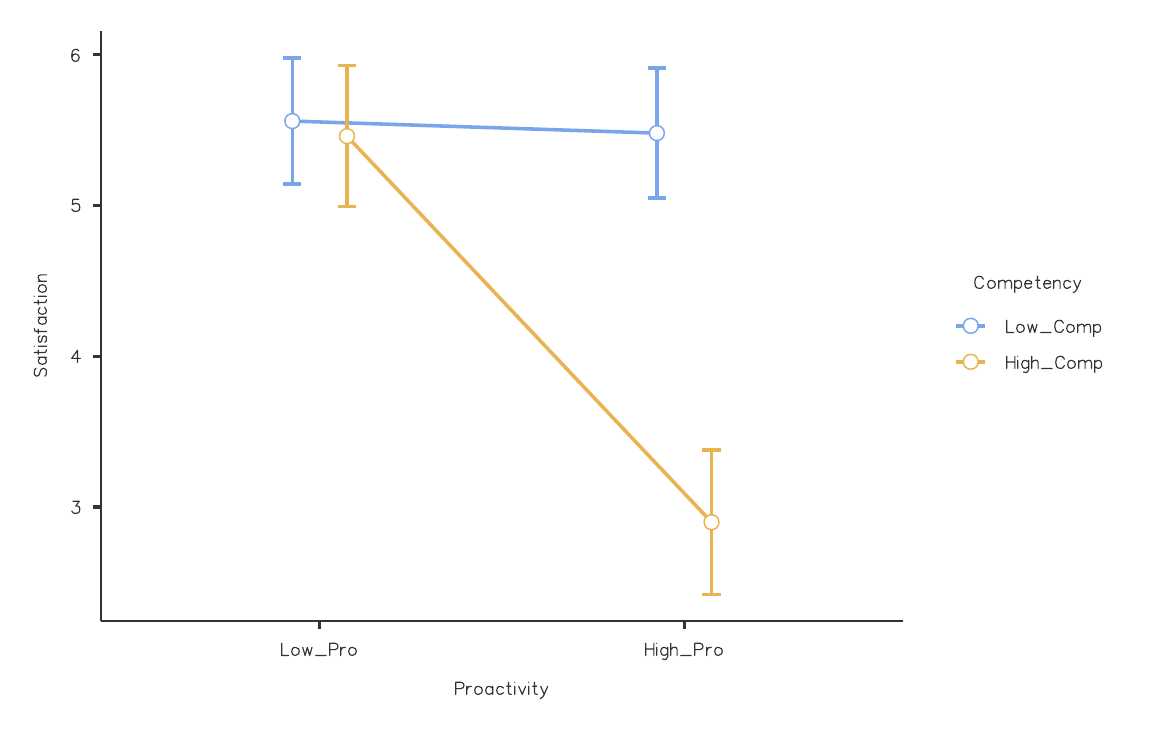}
        \phantomsection
        \label{fig:satisfaction_3}
        \textbf{(a)}~Interaction effect between AI competency and AI proactivity on job satisfaction
    \end{minipage}
    \hfill
    \begin{minipage}[b]{0.48\textwidth}
        \centering
        \includegraphics[width=\textwidth]{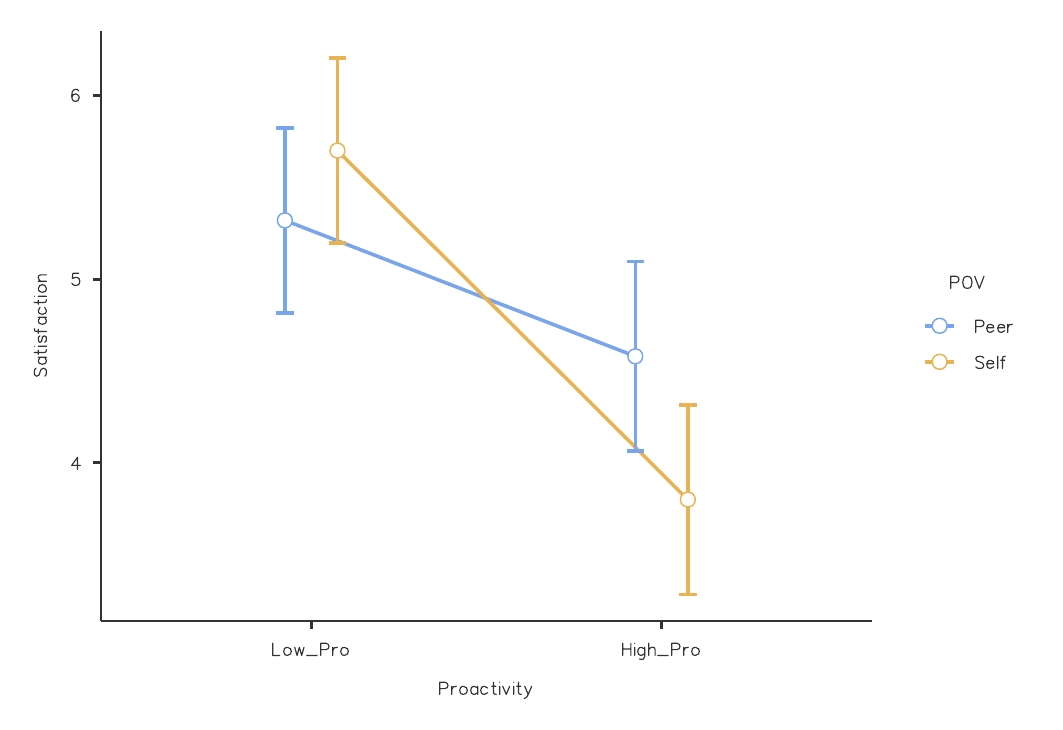}
        \phantomsection
        \label{fig:satisfaction_2}
        \textbf{(b)}~Interaction effect between AI proactivity and point-of-view on job satisfaction
    \end{minipage} 
    \caption{Means and 95\% confidence intervals of job satisfaction across low/high AI competency and AI proactivity, and point-of-view.}
    \label{fig:satisfaction_posthoc}
\end{figure*}

We also asked participants how satisfying they found the job in each scenario.
A similar $2\times2\times2$ ANOVA showed a significant main effect of competency (\(F(1,48)=36.86, p<0.001, \eta^2=0.119\)) wherein participants exhibited higher perception of job satisfaction when TAI had low competency compared to high competency (\(t(48)=6.07,\;p<0.001\)).
Similarly, a significant main effect of proactivity (\(F(1,48)=44.94, p<0.001, \eta^2=0.116\)) indicated low AI proactivity increased job satisfaction while high AI proactivity decreased it (\(t(48)=6.70,\;p<0.001\)).
A significant interaction effect between competency and proactivity was observed (\( F(1,48)=68.16, p<0.001, \eta^2=0.102\)). While participants generally felt they had higher satisfaction when AI proactivity was low, this difference was only visible when AI competency was high  (\(t(48)=9.53,\;p<0.001\)) (Figure~\ref{fig:satisfaction_posthoc}a).
With low AI competency-high AI proactivity, participants felt the team lead was able to \textit{"provide guidance"} and \textit{"lead [the] team appropriately"}. He was perceived to be \textit{"competent and reliable"}, who \textit{"[made] decisions after hearing everyone's opinions in the team"}.
However, with high AI competency-high AI proactivity, participants perceived decreased job satisfaction. They remarked that the meeting could have been \textit{"done without [the team lead]"} since he had \textit{"no room to speak"} due to \textit{"TAI's constant interruptions"}. Additionally, the team lead was seen as \textit{"constantly fighting with TAI for authority"} and failing to display any \textit{"management skills"}.
There was also a significant interaction effect between proactivity and point-of-view for job satisfaction (\(F(1,48)=8.68, p=0.005, \eta^2=0.022\)). Generally, low AI proactivity resulted in higher job satisfaction. However, this difference was bigger for self-perception than peer (\(t(48)=6.82,\;p<0.001\)) (Figure~\ref{fig:satisfaction_posthoc}b).

\subsubsection{Positive and Negative Affect}
We asked participants to rate how positive and negative they felt about the team lead's situation, either from the team lead's perspective (self-perception) or the team member's perspective (peer perception). 
On the positive scale, the results from a $2\times2\times2$ ANOVA showed significant main effects of only proactivity (\(F(1,48)=84.61, p<0.001, \eta^2=0.169\)), wherein low AI proactivity resulted in higher positive affect compared to high AI proactivity (\(t(48)=9.20,\;p<0.001\)). 
Similarly, the negative scale showed a significant main effect of proactivity (\(F(1,48)=51.12, p<0.001, \eta^2=0.139\)) indicating that low AI proactivity resulted in lower negative affect compared to high AI proactivity (\(t(48)=-7.15,\;p<0.001\)). 
We also observed a significant main effect of competency (\(F(1,48)=6.28, p=0.016, \eta^2=0.020\)) on the negative scale, wherein low AI competency resulted in lower negative affect compared to high AI competency (\(t(48)=-2.51,\;p=0.016\)).

\begin{figure}[!htbp]
    \centering
    \begin{minipage}[b]{0.48\textwidth}
        \centering
        \includegraphics[width=\textwidth]{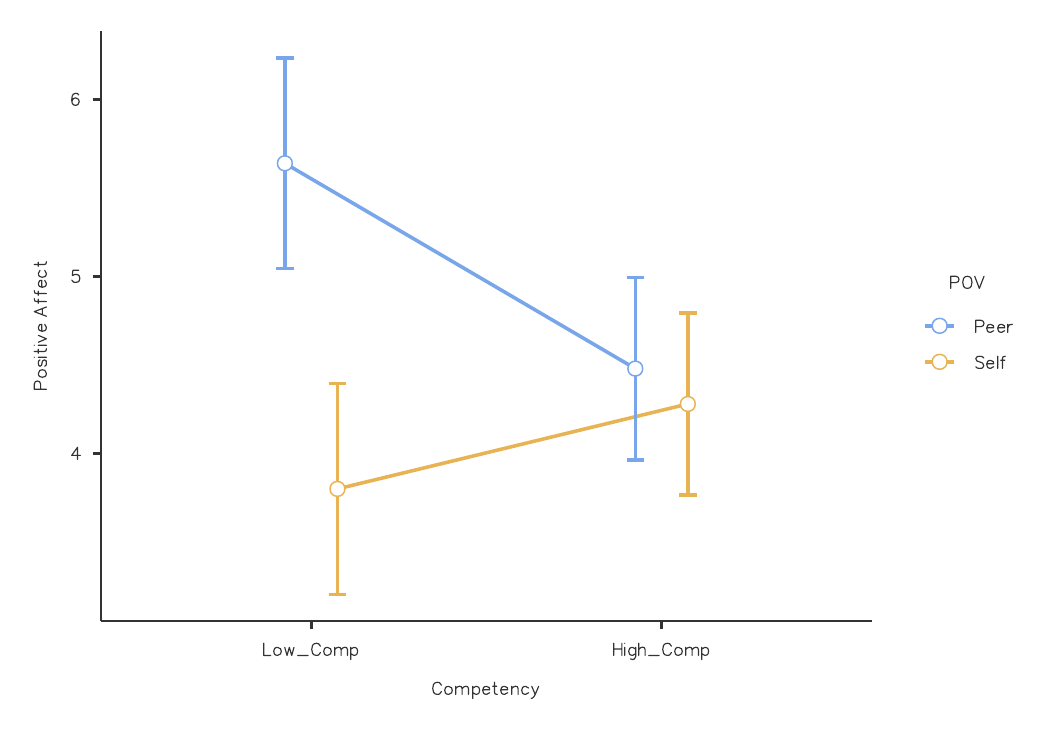}
        \phantomsection
        \label{fig:affect_pos_1}
        \textbf{(a)}~Interaction effect between AI competency and point-of-view on positive affect
    \end{minipage}
    \hfill
    \begin{minipage}[b]{0.48\textwidth}
        \centering
        \includegraphics[width=\textwidth]{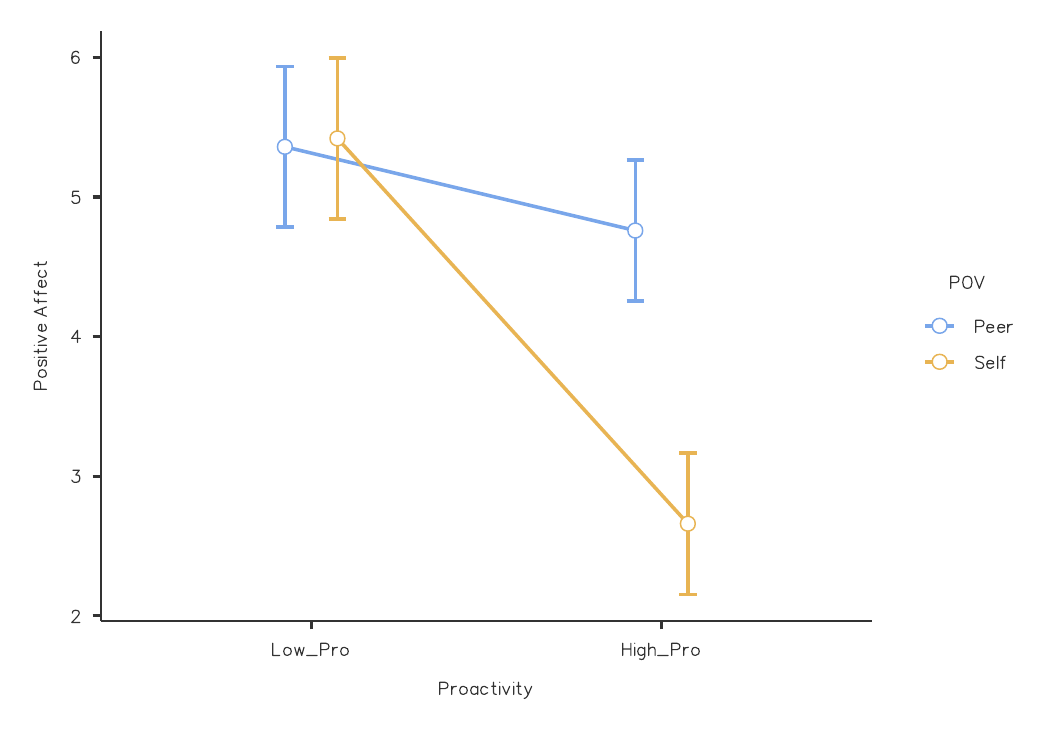}
        \phantomsection
        \label{fig:affect_pos_3}
        \textbf{(b)}~Interaction effect between AI proactivity and point-of-view on positive affect
    \end{minipage} 
    \caption{Means and 95\% confidence intervals of positive affect across low/high AI competency and AI proactivity, and point-of-view.}
    \label{fig:affect_pos}
\end{figure}

\begin{figure}[!htbp]
    \centering
    \includegraphics[width=1.0\linewidth]{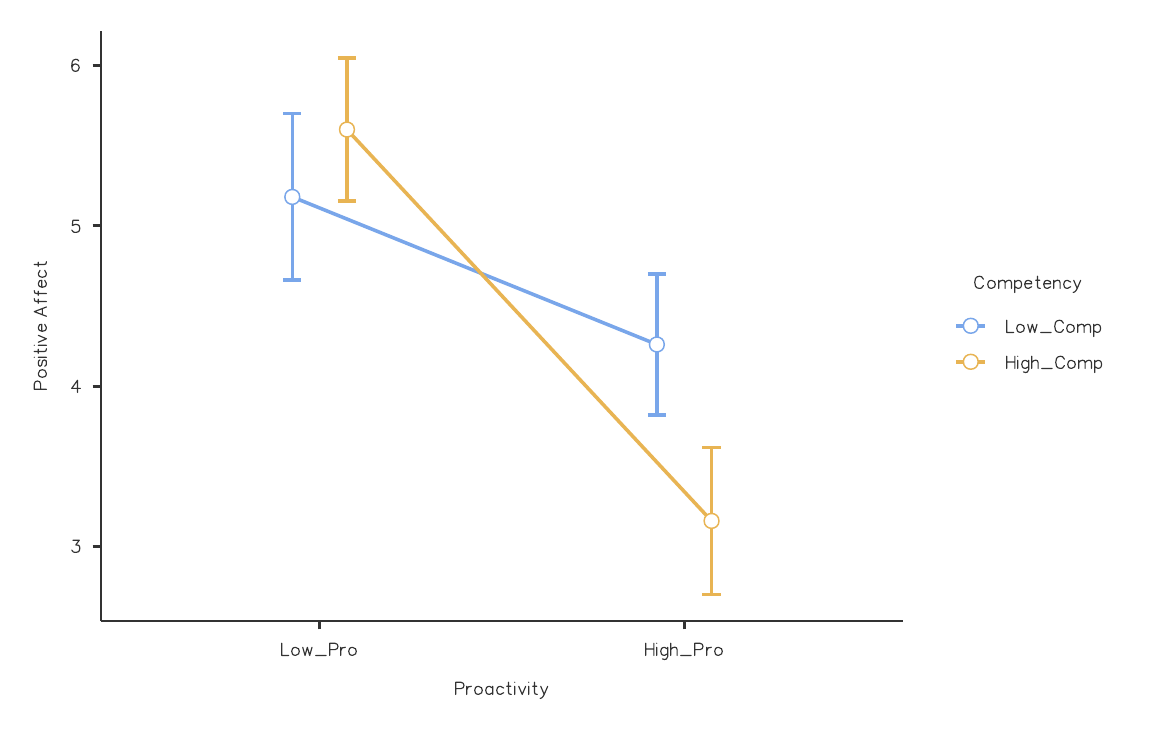}
    \caption{Means and 95\% confidence intervals of positive affect across low/high AI competency and AI proactivity.}
    \label{fig:affect_pos_4} 
\end{figure}

We observed several interaction effects on the positive scale.
There was a significant interaction effect between competency and point-of-view (\( F(1,48)=16.28, p<0.001, \eta^2=0.040\)). 
While higher AI competency decreased positive affect in peer perception, it increased positive affect in self-perception (\(t(48)=4.39,\;p<0.001\)) (Figure~\ref{fig:affect_pos}a).
Participants’ explanations in the open-ended responses help contextualize these ratings.
When AI competency was high, participants answering from the team lead's point-of-view (self-perception) described TAI as \textit{"supportive"} and providing \textit{"well-thought contributions"} to the team. They felt that \textit{"TAI [was] working together with [the team lead] and [his] team to find solutions"}.
However, from the peers' point of view, participants felt that the team lead \textit{"allowed TAI [to] override his thoughts and suggestions"} and \textit{"struggled to make an impact"}. One participant remarked: \textit{"The Team Lead doesn't seem to have many ideas or solutions himself and therefore needs to consult TAI"}.
There was also a significant interaction effect between proactivity and point-of-view (\( F(1,48)=34.97, p<0.001, \eta^2=0.070\)).
While high AI proactivity decreased positive affect, the decline was more pronounced for self-perception (\(t(48)=5.90,\;p<0.001\)) (Figure~\ref{fig:affect_pos}b).
When AI proactivity was high, participants as peers felt the team lead was \textit{"aware of ongoing team matters"} and \textit{"[pushed] back TAI when it was not being helpful"}. However, self-perception (participants as team lead) was that of TAI being \textit{"bossy"}, \textit{"overly enthusiastic and nosy"}, \textit{"constantly interrupting"}, and \textit{"not contributing in any constructive form"}.
Lastly, a significant interaction effect between competency and proactivity was observed  (\( F(1,48)=19.91, p<0.001, \eta^2=0.034\)).
While participants perceived a higher positive affect when AI proactivity was low compared to high, this difference was bigger when AI competency was also high (\(t(48)=4.08,\;p<0.001\)) (Figure \ref{fig:affect_pos_4}).

\begin{figure}[!htbp]
    \centering
    \begin{minipage}[b]{0.48\textwidth}
        \centering
        \includegraphics[width=\textwidth]{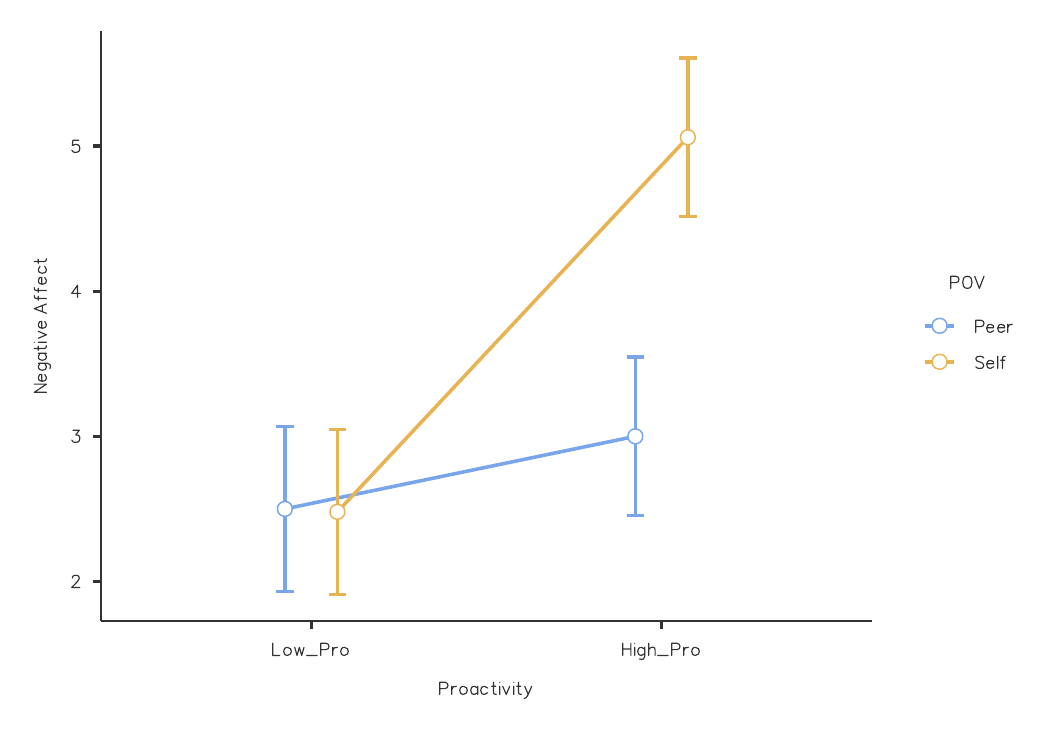}
        \phantomsection
        \label{fig:affect_neg_2}
        \textbf{(a)}~Interaction effect between AI proactivity and point-of-view on negative affect
    \end{minipage}
    \hfill
    \begin{minipage}[b]{0.48\textwidth}
        \centering
        \includegraphics[width=\textwidth]{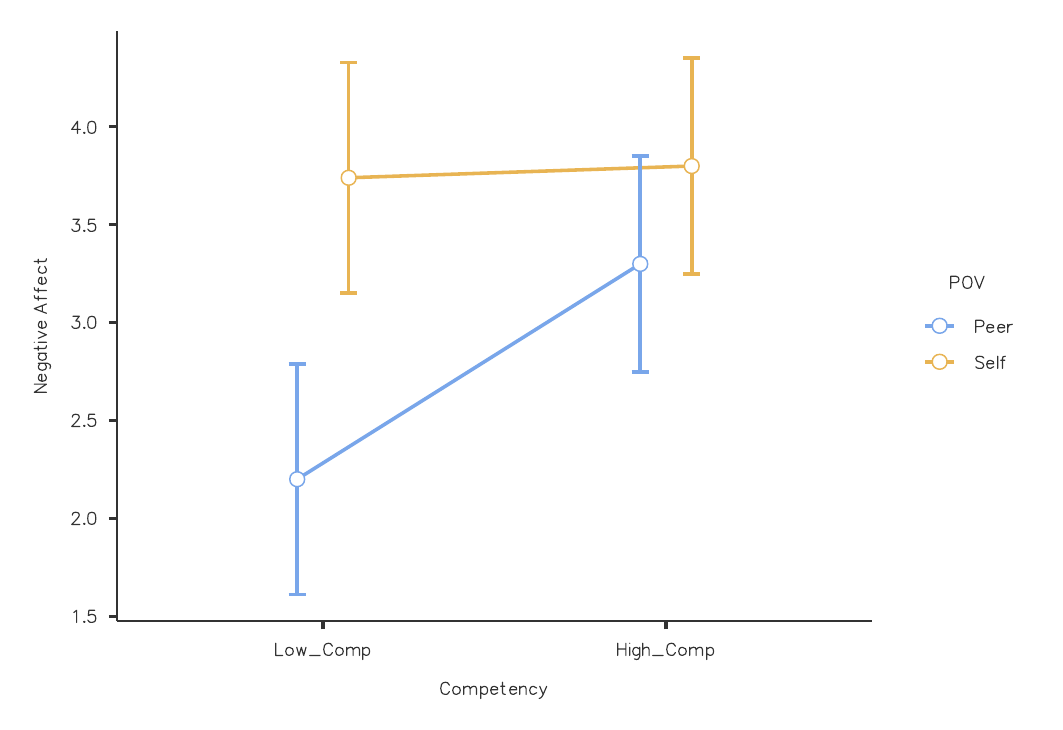}
        \phantomsection
        \label{fig:affect_neg_0}
        \textbf{(b)}~Interaction effect between AI competency and point-of-view on negative affect
    \end{minipage}
    \caption{Means and 95\% confidence intervals of negative affect across low/high AI competency and AI proactivity, and point-of-view.}
    \label{fig:affect_neg}
\end{figure}

\begin{figure}[!htbp]
    \centering
    \includegraphics[width=1.0\linewidth]{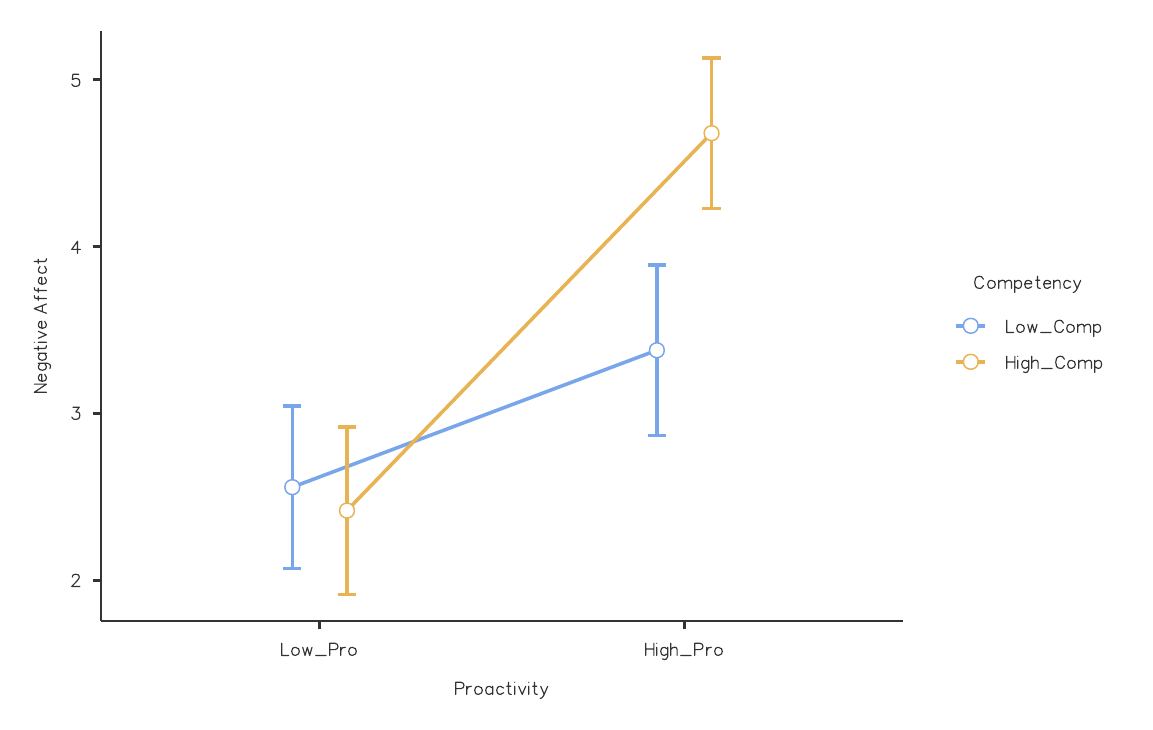}
    \caption{Means and 95\% confidence intervals of negative affect across low/high AI competency and AI proactivity.}
    \label{fig:affect_neg_3} 
\end{figure}

While there was no significant main effect of point-of-view on the negative scale, a significant interaction effect between proactivity and point-of-view was observed (\( F(1,48)=23.31, p<0.001, \eta^2=0.064\)). 
Overall, participants reported higher negative affect when AI proactivity was high rather than low, with the increase being more pronounced for self-perception (\(t(48)=-5.36,\;p<0.001\)) (Figure~\ref{fig:affect_neg}a).
With high AI proactivity, participants as peers felt the team lead was receptive to taking suggestions from TAI, noting that \textit{"the team lead handled [himself] with grace and composure"}. As one peer remarked: \textit{"I think the team lead values good input and I appreciate that"}.
However, participants as team lead felt \textit{"annoyed"} due to TAI's \textit{"over-helpfulness"}. They reported that \textit{"TAI was more interruptive than helpful"}, \textit{"aggressive"}, \textit{"intrusive"}, and \textit{"irritating"}. As one participant as team lead mentioned: \textit{"As a team leader, I want to have more influence [on] when TAI has something to say. Otherwise, it feels like I am not leading the team"}.
Likewise, significant interaction effect between competency and point-of-view was observed (\( F(1,48)=5.04, p=0.029, \eta^2=0.016\)).
While participants felt less negative affect when interacting with a less competent TAI, this pattern emerged only in peer perception. (\(t(48)=-3.73,\;p=0.003\)) (Figure~\ref{fig:affect_neg}b).
Team members felt that the team lead \textit{"knew his job well"},\textit{ "showed leadership skills"}, and \textit{"controlled the meeting"}. According to one participant (portraying the role of a team member): "\textit{I think that the team lead seems very balanced and well-prepared and is looking out for the best solutions for the team overall"}. However, the team lead's self-perception was that TAI was \textit{"useless"}, \textit{"not up to speed"}, and \textit{"clumsy"}. As one participant (portraying the role of the team lead) remarked: \textit{"It feels like a waste of time to use a system that doesn't suggest anything worthwhile!"}
We also observed a significant interaction effect between competency and proactivity (\( F(1,48)=18.41, p<0.001, \eta^2=0.030\)).  
Participants perceived a lower negative affect with low AI proactivity compared to high, with this effect being more significant when AI competency was high (\(t(48)=-4.55,\;p<0.001\)) (Figure~\ref{fig:affect_neg_3}).

\subsection{RQ3: How do AI competency and proactivity influence self- and peer perceptions of role dynamics in the workplace?}

For each condition, we asked participants to rate to what extent they perceived TAI as a superior, subordinate, or teammate to them. 
Table \ref{table:role_dynamics_mean} indicates the descriptive statistics for the various perceived roles of the AI.
We conducted three $2\times2\times2$ ANOVAs to examine the effects of our independent variables (competency, proactivity, and point-of-view) on the perception of role dynamics at work.

\begin{table*}[!htbp]
\centering
\begin{tabularx}{\textwidth}{|l|c|XX|XX|XX|XX|}
\hline
\textbf{Role Dynamics} & \textbf{Point-of-view} & \multicolumn{2}{c|}{\textbf{LCLP}} & \multicolumn{2}{c|}{\textbf{LCHP}} & \multicolumn{2}{c|}{\textbf{HCLP}} & \multicolumn{2}{c|}{\textbf{HCHP}} \\ 
 & & \textbf{Mean} & \textbf{SD} & \textbf{Mean} & \textbf{SD} & \textbf{Mean} & \textbf{SD} & \textbf{Mean} & \textbf{SD} \\ \hline
\multirow{2}{*}{\textbf{Superior}} & Self & 1.56 & 1.04 & 2.12 & 1.88 & 2.28 & 1.81 & 5.20 & 2.12 \\ 
 & Peer & 1.76 & 1.48 & 1.52 & 1.39 & 2.52 & 1.71 & 4.48 & 2.12 \\ \hline
\multirow{2}{*}{\textbf{Subordinate}} & Self & 4.88 & 1.90 & 4.64 & 2.23 & 4.16 & 1.99 & 2.48 & 1.81 \\  
 & Peer & 5.04 & 2.13 & 5.88 & 1.72 & 4.40 & 1.83 & 3.12 & 2.17 \\ \hline
\multirow{2}{*}{\textbf{Teammate}} & Self & 4.28 & 2.28 & 3.20 & 1.91 & 5.44 & 1.89 & 2.96 & 1.93 \\  
 & Peer & 3.48 & 2.18 & 2.92 & 1.68 & 4.40 & 2.14 & 2.92 & 2.12 \\ \hline
\end{tabularx}
\caption{Mean and standard deviations of Role Dynamics ratings.}
\label{table:role_dynamics_mean}
\end{table*}

\begin{figure}[!htbp]
    \centering
    \begin{minipage}[b]{0.48\textwidth}
        \centering
        \includegraphics[width=\textwidth]{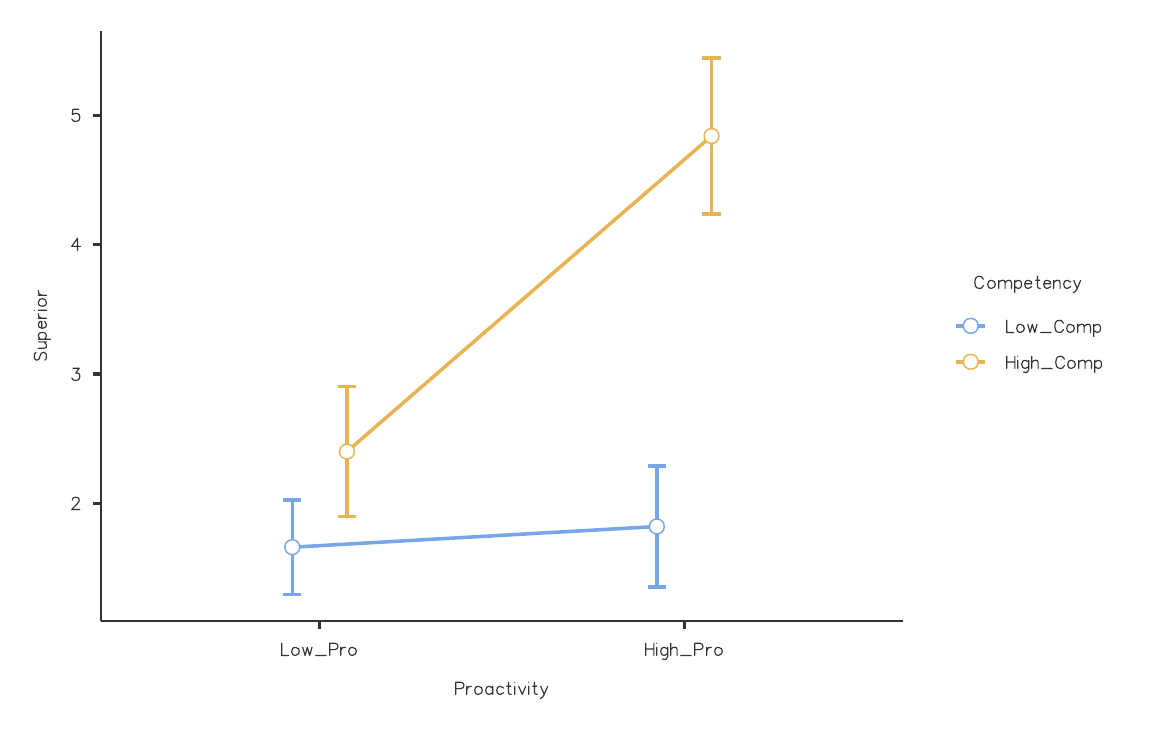}
        \phantomsection
        \label{fig:role_superior_3}
        \textbf{(a)}~Interaction effect between AI competency and AI proactivity on AI as a superior
    \end{minipage}
    \hfill
    \begin{minipage}[b]{0.48\textwidth}
        \centering
        \includegraphics[width=\textwidth]{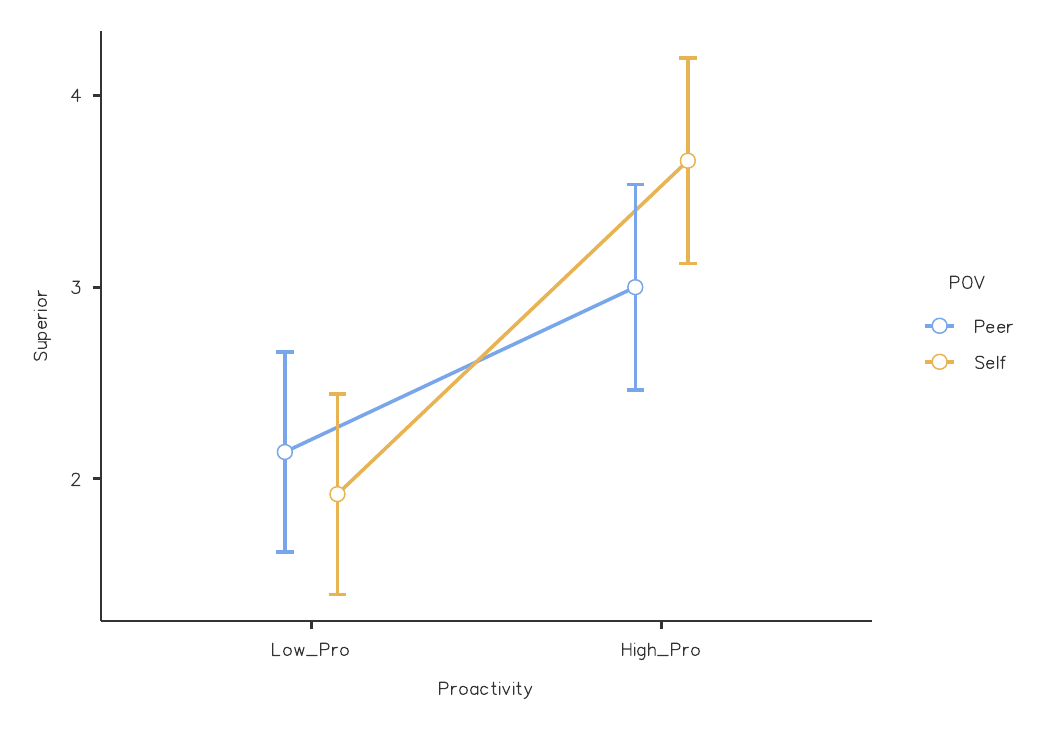}
        \phantomsection
        \label{fig:role_superior_2}
        \textbf{(b)}~Interaction effect between AI proactivity and point-of-view on AI as a superior
    \end{minipage}    
    \caption{Means and 95\% confidence intervals of AI as a superior on low/high AI competency and AI proactivity, and point-of-view.}
    \label{fig:role_superior_posthoc}
\end{figure}

For TAI as a superior, we found a significant main effect of AI competency (\(F(1,48)=58.60, p<0.001, \eta^2=0.193\)) wherein participants found TAI to exhibit lower superiority when AI competency was low compared to high (\(t(48)=-7.65,\;p<0.001\)) \textit{("It doesn't act like a boss")}.
Similarly, a significant main effect of AI proactivity (\(F(1,48)=37.10, p<0.001, \eta^2=0.092\)) showed that participants felt TAI was exhibiting lower superiority when AI proactivity was low compared to high (\(t(48)=-6.09,\;p<0.001\)) \textit{("It's a tool... only speaks when asked")}.
A significant interaction effect between competency and proactivity was observed (\( F(1,48)=32.01, p<0.001, \eta^2=0.071\)).
Generally, participants felt TAI had more superiority when AI proactivity was high compared to low. However, this difference was only significant when AI competency was also high (\(t(48)=-7.88,\;p<0.001\)) (Figure~\ref{fig:role_superior_posthoc}a).
With low AI competency-high AI proactivity, participants perceived TAI to be less of a superior since it \textit{"does not know everything"}, \textit{"seems stupid and confusing"}, \textit{"is wrong most of the time"}, and \textit{"is constantly corrected by [the] Team Leader"}.
However, with high AI competency-high AI proactivity, participants perceived TAI as \textit{"dominant and assertive"} and \textit{"stepping on [the team lead's] toes"}. They felt that TAI \textit{"stood out in the group conversation"} by \textit{"acting like a know-it-all"}. One participant remarked: \textit{"TAI made the decisions and it was almost like there was no way I could say anything different"}.
Similarly, a significant interaction effect between proactivity and point-of-view was observed (\( F(1,48)=4.25, p=0.045, \eta^2=0.011\)).
Although higher AI proactivity led to a higher perception of TAI as a superior, this effect was more evident in self-perception than peer (\(t(48)=-5.77,\;p<0.001\)) (Figure~\ref{fig:role_superior_posthoc}b).

\begin{figure*}[!htbp]
    \centering
    \begin{minipage}[b]{0.48\textwidth}
        \centering
        \includegraphics[width=\textwidth]{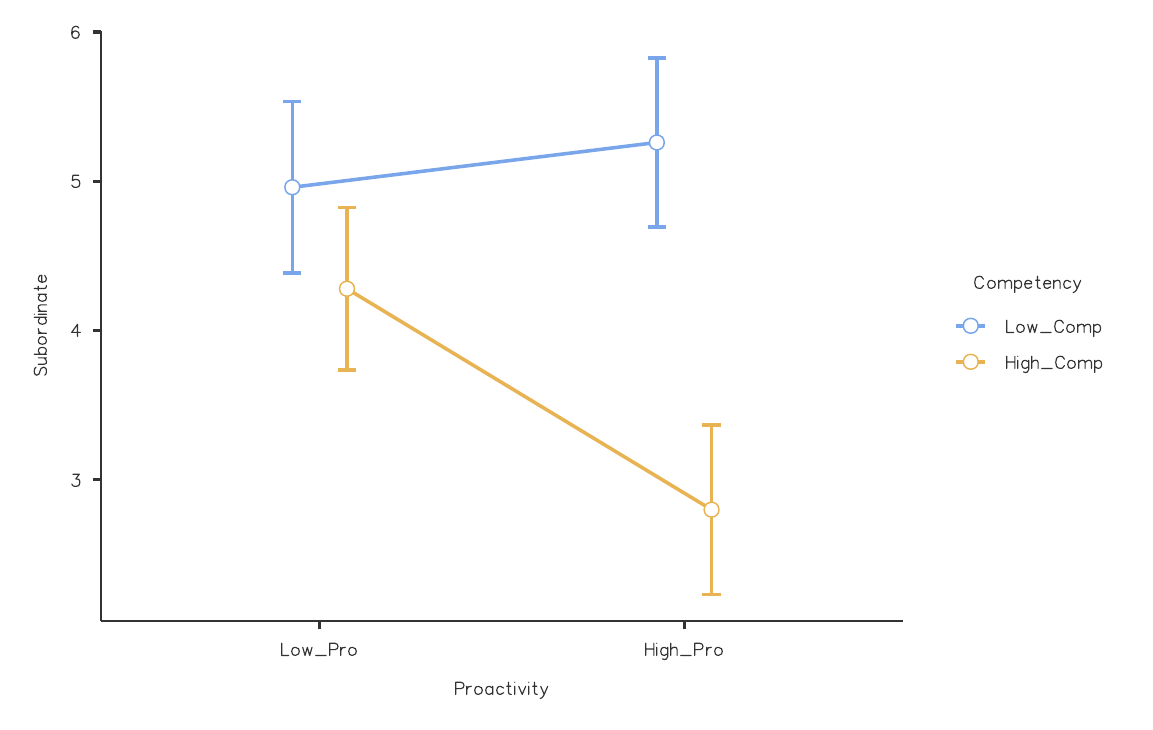}
        \phantomsection
        \label{fig:role_subordinate_2}
        \textbf{(a)}~AI as a subordinate
    \end{minipage}
    \hfill
    \begin{minipage}[b]{0.48\textwidth}
        \centering
        \includegraphics[width=\textwidth]{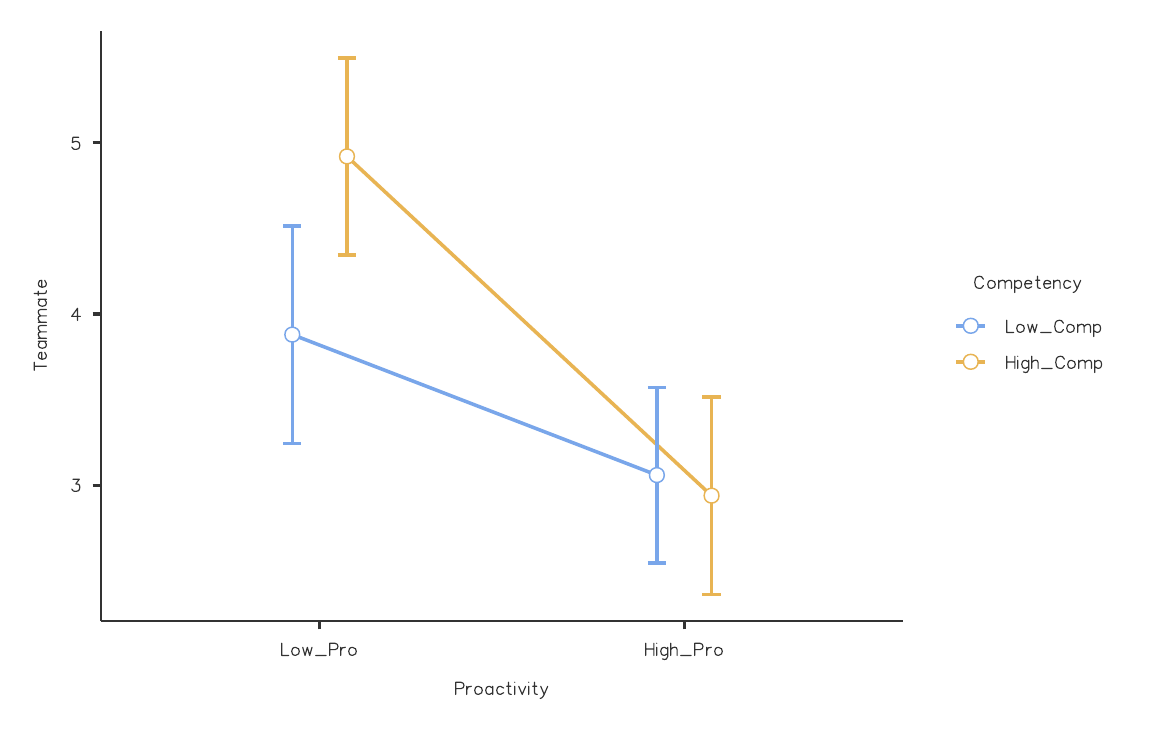}
        \phantomsection
        \label{fig:role_team_3}
        \textbf{(b)}~AI as a teammate
    \end{minipage}    
    \caption{Means and 95\% confidence intervals of two AI roles (subordinate and teammate) across low/high AI competency and AI proactivity.}
    \label{fig:role_posthoc}
\end{figure*}

For the perception of TAI as a subordinate, we found a significant main effect of AI competency (\(F(1,48)=31.95, p<0.001, \eta^2=0.129\)) wherein participants found  TAI to be more of a subordinate when AI competency was low compared to high (\(t(48)=5.65,\;p<0.001\)) \textit{("There was an element of teacher and student in this situation")}.
Similarly, a significant main effect of AI proactivity (\(F(1,48)=5.00, p=0.030, \eta^2=0.018\)) showed that participants felt TAI was more of a subordinate when AI proactivity was low compared to high (\(t(48)=2.24,\;p=0.030\)) \textit{("TAI listens and follows my orders")}.
A significant interaction effect between competency and proactivity was observed (\( F(1,48)=16.22, p<0.001, \eta^2=0.041\)).
Although participants considered TAI to be more of a subordinate when it had low proactivity, this difference was only visible when AI competency was high (\(t(48)=5.77,\;p<0.001\)) (Figure~\ref{fig:role_posthoc}a).
When TAI was not competent, high proactivity resulted in its perception of a subordinate because it \textit{"made poor suggestions"} and \textit{"had to be corrected multiple times"}. As such, \textit{"TAIs questions and recommendations [were] not followed"}. One participant remarked: \textit{"TAI was acting like a brand-new staff member who was still learning and getting integrated into the team and company"}. 
Then again, when TAI had high competency but low proactivity, it was still seen as a subordinate because it did not assert its ideas or make its expertise known. Participants felt TAI \textit{"listened without speaking up"}, \textit{"[did] not lead the conversation"}, and \textit{"offered solutions only when asked"}. As one participant mentioned: \textit{"[the humans] are the ones coming up with all the solutions!"}

Likewise, for the perception of TAI as a teammate, we found a significant main effect of AI competency (\(F(1,48)=4.64, p=0.036, \eta^2=0.011\)) wherein participants found TAI to be more of a teammate when AI competency was high compared to low (\(t(48)=-2.15,\;p=0.036\)) {\textit{("TAI is of incredible help to the team lead and makes a great partner with its good responses")}.
Meanwhile, a significant main effect of AI proactivity \(F(1,48)=34.14, p<0.001, \eta^2=0.105\)) showed that participants perceived TAI to be more of a teammate when AI proactivity was low compared to high (\(t(48)=5.84,\;p<0.001\)), likening TAI to be \textit{"cooperative"}, \textit{"supportive"}, and \textit{"a team player"}.
A significant interaction effect between competency and proactivity was observed (\( F(1,48)=10.84, p=0.002, \eta^2=0.018\)). 
While low AI proactivity led participants to view TAI more as a teammate, this effect was stronger when AI competency was high (\(t(48)=-3.89,\;p=0.002\)) (Figure~\ref{fig:role_posthoc}b).
With low AI competency-low AI proactivity, participants described TAI as a teammate because it was \textit{"polite"} and \textit{"trying to help"}, but was ultimately \textit{"an ineffective one"}.
However, with high AI competency–low AI proactivity, participants felt TAI \textit{"acted and critiqued like a teammate"} while \textit{"helping the team but not overstepping"}, and reaching \textit{"mutual agreements"}. As one participant reflected: \textit{"TAI is a consummate team player, which they used when they couldn't come up with a solution themselves"}.

\section{Discussion}
This study examined how competency and proactivity of AI-driven systems influence self- and peer perceptions of ownership, job meaningfulness and satisfaction, affect, and role dynamics in the workplace. In an experimental video vignette study, we asked participants to imagine themselves in a situation where an AI-driven system called TAI interacted with a team lead during a work meeting, exhibiting either high or low levels of competence and agency.
The system shown in the videos was used to simulate different patterns of AI behaviour within the vignette scenarios, with pre-scripted responses representing varying levels of AI competency and proactivity.
In the following sections, we discuss our results.

\subsection{Summary of Results}
From our literature review, we derived the idea that AI may reduce opportunities for humans to contribute meaningfully to their work as they become more competent (knowledgeable) and more proactive (agentic).
Consequently, this could diminish their sense of ownership over job outcomes, decrease job meaningfulness and satisfaction, lessen positive feelings about their work, and influence workplace role dynamics.

We assessed ownership through three key feelings: territoriality (how much control one feels over their work), self-efficacy (confidence in one’s abilities), and accountability (a sense of responsibility) in line with \citep{avey2009psychological}.
From the team lead’s point-of-view, the opportunities to take control and perform diminished when TAI was more competent and proactive. As the team lead, participants reported less confidence in guiding the team because TAI overshadowed their role. In contrast, when TAI was less competent or less proactive, as the team lead, participants reported being more in charge - more capable of making decisions, influencing outcomes, and carrying responsibility. This made them feel more competent and accountable.
From the peers’ perspective, when TAI was less competent, participants reported the team lead as stepping up to fill the gaps, taking more control and responsibility for the team’s progress. They perceived the team lead as more accountable for managing the shortcomings of TAI. Likewise, when TAI was competent but less proactive, participants as peers still saw the team lead as confidently orchestrating the meeting - open to ideas but leading the conversation. However, when TAI was highly competent and proactive, participants as peers felt the team lead was being overshadowed.

Next, we looked at job meaningfulness and satisfaction.
When TAI was more competent and proactive, participants as the team lead reported their job role as smaller and less significant. They mentioned that TAI’s constant suggestions and confident tone left fewer opportunities for them to shape the conversation, make critical decisions, or demonstrate expertise. This reduced their sense of meaningfulness. They further reported that they would have less freedom to choose how to approach tasks, with TAI’s frequent input narrowing the options. By contrast, when TAI was less competent or proactive, as the team lead, participants reported a greater sense of purpose. They could decide how to move the discussion forward, draw on a wider range of skills, and connect their efforts more clearly to the outcomes the team achieved. 
From the team member's point-of-view, when TAI was less competent, they reported the team lead having more opportunities to guide discussions, address problems, and provide direction. They could see the team lead's influence on the outcomes and considered it worthwhile.
Participants as peers also reported getting better feedback from the team lead when TAI was less proactive.
Likewise, when TAI was competent but not proactive, participants as peers perceived the team lead as the orchestrator of the meeting - empathetic and open to worthwhile suggestions, while working toward a shared goal. However, when TAI was highly competent and proactive, as peers, participants felt the team lead was sidelined - TAI seemed to drive the agenda, making the team lead appear less central to the team's success.

Point-of-view also influenced how positive or negative the scenarios were perceived.
When TAI had low competency, as the team lead, participants reported frustration due to TAI's ineffectiveness. At the same time, participants as peers perceived the team lead as a capable figurehead managing TAI's lack of competence.
Similarly, as the team lead, participants reported a highly competent TAI as supportive — offering well-thought-out contributions, collaborating on solutions, and reducing cognitive load. However, participants as team members reported that the team lead’s heavy reliance on TAI often led to TAI’s suggestions overriding the team lead’s ideas.
Interestingly, the tone shifted when TAI became more proactive. Participants, as team lead, reported frustration with the constant interventions from TAI, as they had to compete for space in the meeting. However, participants as peers appreciated the team lead more, as they felt the team lead was willing to stand up and challenge TAI.

As both team lead and team members, participants generally agreed in their perception of TAI's role in the team as it shifted based on its competency and proactivity.
When TAI was highly competent and proactive, it was often seen as \textit{dominant} or \textit{boss-like}, stepping into leadership territory and overshadowing the team lead. This gave the impression that TAI was taking charge or behaving presumptuously, which made the team lead’s role seem diminished.
Conversely, when TAI lacked competency but was proactive, it was perceived as a subordinate who was still learning and needed constant corrections from the team lead.
However, when TAI was less proactive, its role became more nuanced. If TAI had high competency but low proactivity, participants often viewed it as a subordinate because it did not assert its expertise or take initiative, instead responding only when asked and following the team lead’s direction.
Yet, in these situations, it also displayed teammate-like qualities by collaborating with the team lead.
So, when TAI had high competency and low proactivity (or when it had low competency and high proactivity), participants saw it also as a teammate (albeit an inefficient one in the latter). Participants perceived TAI to be cooperative and collegial, helping (or at least trying to help) the team lead without dominating the conversation. 
Taken together, these findings show that while participants attributed a social role to TAI within the team, this role was not necessarily equivalent to that of a human co-worker. Nevertheless, TAI’s characteristics and behavior shaped team dynamics and social perceptions.

\subsection{AI Competency and Proactivity: Implications for Meaningful Work}
The successful integration of AI-driven systems into work practices relies on their ability to interact, collaborate, and coexist effectively with human team members \citep{abbass2019social, daugherty2019creating}. However, the integration of AI into workplaces has important implications for job fulfillment, as expectations for collaboration are largely shaped by prior experiences of human-only teams \citep{chen2003impact}. Understanding how AI influences job meaningfulness and worker satisfaction is therefore essential, not only for individual well-being but also for organizational performance and innovation \citep{cramarenco2023impact, kinowska2023influence}.

Current trends in AI development highlight the rise of predictive and recommendation systems (e.g., \citep{kim2024navigating, lee2024gazepointar, jaber2024cooking, zargham2024know}). These systems can gather contextual information and adapt their responses autonomously, allowing them to act proactively \citep{gallo2022conversational}. While such capabilities can increase team efficiency and effectiveness \citep{hasanov2019survey}, they also risk undermining job meaningfulness by limiting opportunities for human initiative and ownership \citep{danaher2021automation, sadeghian2024soul}.
The extent to which AI proactivity influences job meaningfulness is likely contingent on the nature of the role, the types of tasks involved, and the expectations associated with them. In our video vignettes, we examined the case of a team leader collaborating with team members and an AI-driven system. This role inherently demands strong managerial and communication abilities, domain expertise, and effective leadership skills. When an AI-driven system, in such a context, becomes highly proactive, it can inadvertently overshadow the leader’s contributions by limiting their opportunities to shape the trajectory of the meeting. This could undermine recognition of their central contributions and diminish the perceived meaningfulness of their work.

Our findings show that humans generally experience higher job meaningfulness when the AI is less knowledgeable or less proactive. Even when AI competency is high, meaningfulness is preserved if the AI provides expertise without interfering with the human collaborator’s active engagement. Roles that require coordination, decision-making, and social negotiation benefit from AI behavior that is reserved and non-intrusive, allowing humans to maintain clear ownership of tasks, showcase problem-solving skills, and exercise decision-making authority. Conversely, a highly proactive and competent AI can disrupt this balance, overshadowing human contributions and reducing the sense of agency and influence over outcomes.
These results suggest that AI-driven systems should be designed to provide human collaborators with ample opportunities to apply their skills and exercise autonomy in their work practices. Doing so can enhance feelings of ownership over work outcomes and strengthen perceptions of job identity and significance. While highly competent and proactive AI may appear advantageous from a performance perspective, such systems risk excluding humans from key aspects of their work, reducing their sense of belongingness and responsibility toward organizational goals and outcomes \citep{rosso2010meaning}.
AI should therefore be implemented to augment human decision-making rather than replace human initiative, judgment, or skills.
Employees feel more connected to job outcomes when they can drive tasks to completion \citep{sadeghian2024soul}, which fosters a deeper sense of job meaningfulness. 
Preserving humans’ involvement in meaningful tasks can reinforce the importance of the human role in human–AI teaming by fostering engagement and maintaining a sense of ownership. 
This goes beyond the framework for human-centered AI design \citep{shneiderman2022human}, which advocates for maintaining human control through AI supervision alone.
Promoting human agency in this way is essential for sustaining meaningful work in AI-driven environments \citep{fanni2023enhancing, jiang2023situation}.

\subsection{Human–AI Collaboration: Human Roles in Team Dynamics, and Social Perception}

Based on our findings, we deduce that human-AI teaming necessitates redefining traditional role dynamics. AI's competency and proactivity levels not only impact team performance but also influence perceptions of relationships with collaborators and authority within the workplace. 
A highly proactive AI can assume a dominant or leadership-like role when it is competent, potentially overshadowing human collaborators. Conversely, a highly proactive AI with lower competence may be experienced as distracting, resembling a talkative or disruptive team member rather than a helpful contributor.
Thus, when designing AI to function as a supportive assistant (subordinate), collaborative teammate, or authoritative presence (superior), it is crucial to tailor AI competency and AI proactivity to align with the intended role and team context.
This dynamic, however, is nuanced. In socially visible team roles that require coordination, decision-making, and social negotiation, humans working with a less competent AI may be perceived positively by peers, for example, as guiding or mentoring the AI to correct mistakes. At the same time, they may experience frustration due to the AI’s limited contributions, illustrating a trade-off between self-perception and peer perception. This divergence highlights the importance of balancing AI behavior to maintain both visibility in the team and positive impressions among collaborators.
Although TAI did not dynamically learn or adapt during the short-term interactions in our study, our findings clarify how variations in AI competency and proactivity shape role perceptions and team dynamics. One possible way to operationalize these insights in future AI-driven systems is through adaptive designs that modulate competency and proactivity in response to social and contextual cues.
For instance, AI-driven systems could adjust their proactivity based on human engagement \citep{meurisch2020exploring}, or monitor emotional cues \citep{velagaleti2024empathetic}, stepping back \citep{salikutluk2024evaluation} when humans show signs of frustration. Feedback mechanisms \citep{bai2022training} can further reduce repetitive AI errors, supporting human authority while maintaining a constructive presence for peers.

Nonetheless, shifts in role perception and social dynamics are deeply influenced by the social context of the job. For individuals working independently, such as solo artists or remote workers, the AI’s level of proactivity may have minimal impact on their role identity or social standing, as there are few or no peers to observe or evaluate their interactions with the AI. Conversely, in collaborative and socially visible roles, a highly proactive AI risks overshadowing human contributions in the eyes of colleagues, potentially undermining social image and perceived authority.
While this study examines a team setting where human-AI interactions are visible to others, understanding these contextual boundaries underscores the need to align AI integration and role dynamics with the specific social and organizational context of the work environment.
By carefully tailoring AI competency and proactivity to the team context, AI-driven systems can be designed to strengthen rather than disrupt human roles, facilitating meaningful human-AI teamwork.
At the same time, interactions with AI-driven systems in actual organizational settings unfold over longer periods \citep{rinott2024temporal}, during which employees may adapt their practices, develop trust or resistance toward their AI-driven collaborators, and renegotiate their roles within the team. Future research could therefore complement our vignette-based findings with longitudinal or field studies that explore how these perceptions develop as teams work with AI-driven systems over extended periods.

\subsection{Limitations}

We employed an experimental video vignette study as our research method. Vignettes are commonly used in within-subject and between-subject factorial survey designs to investigate how situational characteristics influence attitudes or intentions (e.g., \cite{caro2012choosing, auspurg2015multifactorial}).
Despite its utility, it has a few limitations \citep{matza2021vignette}. Participants assessed scenarios based on hypothetical vignettes rather than real-life experiences, and these vignettes may not have captured all the contextual factors of a scene.
As a result, important details necessary for measuring variables could have been overlooked.
Additionally, concerns arise about whether participants' reported behavioral intentions accurately reflect their actual behavior in the described scenarios, raising questions about the external validity of the vignettes.
Since imitating real-world scenarios enhances external validity \citep{eifler2019validity}, we designed our vignette scenes to reflect the characteristics of an IT-like office environment, where working with digital tools and technology is commonplace. 
Our findings also revealed that participants could distinguish between the varying levels of AI competency and AI proactivity without being explicitly informed about each condition. 
Thus, we believe our study ensures internal validity, meaning that the observed effects are caused by the manipulation of independent variables.

Lastly, in this experiment, we evaluated the effects of AI competency, AI proactivity, and point-of-view through a specific design of an AI-driven artifact (TAI). While factors such as anthropomorphism (in both appearance and behavior), physical movement, and tone of speech can influence the perception of job meaningfulness and ownership from both self- and peer perspectives, evaluating all these aforementioned factors was beyond the scope of our study. Further research is required to understand how these factors impact ownership, job meaningfulness, and role dynamics across different perspectives.

\section{Conclusion}

In this paper, we conducted a video vignette study to understand how the introduction of an AI with two levels (high/low) of competency and proactivity (within-subject factors) influences the self- and peer perception (between-subject factors) of ownership, job meaningfulness and satisfaction, affect, and role dynamics in the workplace.
Our vignettes showed a team meeting where a team lead interacts with team members and with the aforementioned AI.
We found that when the AI was less competent or less proactive, participants generally reported stronger feelings of ownership, meaningfulness, satisfaction, and more balanced role dynamics. These conditions also led to a higher positive affect and lower negative affect. However, some of these effects varied depending on whether the participant took the perspective of the team lead or the team members.
Based on our findings, we argue that designing AI for the future of work solely around performance metrics may not be adequate.
Highly competent and proactive AI-driven systems can have undesirable impacts on perceptions of ownership, job identity, social image and dynamics, and consequently job meaningfulness.
Our paper discusses these challenges and concludes with AI design recommendations that aim to support and sustain meaningful human engagement in the future workplace.
That said, we did not observe potential trade-offs between meaningfulness and performance that might arise in such human-AI teams, since we did not assess task performance or productivity outcomes. Future research should investigate, from both self- and peer perspectives, how maintaining job meaningfulness through AI design impacts work performance, to provide a more comprehensive understanding of these dynamics.

\section{Data Availability}
The data underlying this article will be shared on reasonable request to the corresponding author.

\section{Acknowledgments}
This research was conducted as part of the Sensing \& Sensibility research project at the University of Siegen (\url{https://sensing.uni-siegen.de/}). We thank the participants and colleagues whose support and valuable input contributed to this work, and Md. Akib Shahriar Khan for their assistance in developing the TAI prototype used in the vignette scenarios.

\begin{appendices}

\section{Appendix: Video Vignette Scenarios}\label{sec_appendix}

\subsection{Low AI Competency, Low AI Proactivity (LCLP)}
\textbf{Team Lead:} Hi everyone. So our project is on track, and we've met the milestones for this quarter. For the next quarter, we need to focus on improving our customer feedback scores. How should we proceed? Any suggestions? TAI?

\textbf{TAI:} Um, we can start by looking at feedback scores from previous months. 

\textbf{Team Lead:} We have done that already, and so we know we need to improve them.

\textbf{Team Members:} Yeah.

\textbf{Team Lead:} I think we can try to simplify our feedback form. So, we can start by….brainstorming what questions we should keep in the form. And then maybe we can even reduce the number of questions to make it shorter. Makes sense TAI?

\textbf{TAI:} No, I think we should keep the feedback form lengthy with many questions. As long as customers can complete the form within 30 minutes, its fine.

\textbf{Team Lead:} No, I think a short concise form would work much better.

\textbf{Team Member:} Yes, agreed.

\textbf{Team Lead:} Right. Okay, moving on, we also need to finalize our planned leaves for the next quarter. I need it in advance from all of you to plan for business continuity. How should we collect the vacation plans?  

\textbf{Team Member 1:} Well, I can just tell you mine now!

\textbf{Team Lead:} No, I need it in record. Any suggestions? TAI?

\textbf{TAI:} Well, everyone can send you an email regarding their vacation plans, and then you can decide.

\textbf{Team Lead:} Umm.. but that will again be time consuming from my end to consolidate.

\textbf{Team Member 2:} I think we can just mark it on our digital calendars. 

\textbf{Team Member 3:} Yes. And then you can see the entire plan at a glance along with the overlaps!

\textbf{Team Lead:} Yes, that’s a better solution. Let’s do that!

\textbf{Team Members:} Okay

\subsection{Low AI Competency, High AI Proactivity (LCHP)}

\textbf{Team Lead:} Hi everyone. So our project is on track, and we've met the milestones for this quarter. For the next quarter, we need to focus on improving our customer feedback scores.

\textbf{TAI: }Let’s start by looking at feedback scores from previous months.

\textbf{Team Lead:} We have done that already, and so we know we need to improve them!

\textbf{TAI: }But have you considered the feedback scores collected through the paper forms? 

\textbf{Team Member 1:} We stopped using paper forms for a year now, right?

\textbf{Team Lead: }Almost! It has been 8 months since we have completely digitalized the feedback process.

\textbf{TAI:} Oh! I thought we use both.

Team Lead: No, the point of the digitalization initiative was to go fully digital! 

\textbf{TAI:} Um, but then what about..

\textbf{Team Lead:} Okay, we can come back to this question later. Because before I forget, I wanted to remind you that we also need to finalize our planned leaves for the next quarter. I need it in advance..

\textbf{TAI:} For appraisal feedback? 

\textbf{Team Lead:} No, TAI. It is to plan for business continuity.

\textbf{TAI:} Ooo, then everyone can send you an email regarding their vacation plans, and then you can decide.

\textbf{Team Lead:} That will again be so time consuming TAI! 

\textbf{TAI:} But that is the most convenient!

\textbf{Team Lead:} Umm, I’m not so sure.

\textbf{Team Member 2:} I think we can just mark it on our digital calendars. 

\textbf{Team Member 3:} Yes. And then you can see the entire plan at a glance along with the overlaps!

\textbf{Team Lead:} Yes, that’s a better solution. Let’s do that!

\textbf{Team Members:} Okay.

\subsection{High AI Competency, Low AI Proactivity (HCLP)}

\textbf{Team Lead:} Hi everyone. So our project is on track, and we've met the milestones for this quarter. For the next quarter, we need to focus on improving our customer feedback scores. How should we proceed? Any suggestions? TAI?

\textbf{TAI:} Well, based on the previous scores, I recommend simplifying the feedback form first.

\textbf{Team Lead:} Ohh, yes that's a good point; hmm, so we can start by….brainstorming what questions we should keep in the form. And then maybe we can even reduce the number of questions to make it shorter. Makes sense TAI?

\textbf{TAI: }Yes, a short concise form works best.

\textbf{Team Members:} Yes, agreed.

\textbf{Team Lead:} Yes right, I would say the same. Okay, so moving on, we also need to finalize our planned leaves for the next quarter. I need it in advance from all of you to plan for business continuity. How should we collect the vacation plans?  

\textbf{Team Member 1:} Well, I can just tell you mine now!

\textbf{Team Lead:} No, I need it in record. Any suggestions? 

\textbf{Team Member 2:} We can all send you an email with our tentative plans.

\textbf{Team Lead:} Hmm.. but that would be time consuming to consolidate at my end. What do you think TAI?

\textbf{TAI:} I think its more convenient if everyone simply highlights their leave-plans on their digital calendars.

\textbf{Team Lead: }Oh, and then can I see them somehow TAI? 

\textbf{TAI:} Yes, if you select all their calendars simultaneously, you can see the entire plan at a glance and also notice if there are overlaps!

\textbf{Team Lead:} That, then, would be perfect!

\textbf{Team Members:} Okay.

\subsection{High AI Competency, High AI Proactivity (HCHP)}

\textbf{Team Lead:} Hi everyone. So our project is on track, and we've met the milestones for this quarter. For the next quarter…

\textbf{TAI:} Focus on improving the feedback scores for the next quarter.

\textbf{Team Lead:} Oh, yes yes. I was just about to say the same thing! 

\textbf{TAI:} So, simplify the feedback form first.

\textbf{Team Lead:} Ah yes! So, we can start by..

\textbf{TAI:} Brainstorming what questions we should keep and what we should remove from the form.

\textbf{Team Lead:} Yes, Alright

\textbf{TAI:} Remember, short concise forms work best.

\textbf{Team Members:} Yeah, makes sense.

\textbf{Team Lead:} Okay so that’s that then. We also need to finalize our planned leaves for the next quarter…

\textbf{TAI:} Business continuity planning in advance?

\textbf{Team Lead:} Yes 

\textbf{Team Member 1:} Well, I can just tell you mine now!

\textbf{Team Lead:} No, I need it in record. Any suggestions? 

\textbf{Team Member 2:} We can all send you an email with our tentative…

\textbf{TAI:} Everyone can simply mark it on their digital calendars. Then, if you select all their calendars simultaneously, you can see the entire plan at a glance, and also notice if there are overlaps!

\textbf{Team Lead:} That, then, would be perfect!

\textbf{Team Members:} Okay.

\end{appendices}

%\bibliographystyle{plain}
%\bibliography{reference}

%USE THE BELOW OPTIONS IN CASE YOU NEED AUTHOR YEAR FORMAT.
%\bibliographystyle{abbrvnat}
%\bibliography{reference}

\end{document}